\documentclass{article}
\usepackage{graphicx} 
\usepackage{caption}
\usepackage{amsthm}
\usepackage{amssymb}
\usepackage{amsmath}
\usepackage{xcolor}
\usepackage{hyperref}
\usepackage{tikz}
\usetikzlibrary{decorations.pathreplacing,angles,quotes,calc}
\usepackage{fontawesome5}
\usepackage{enumerate}
\usepackage{mathabx} 
\usepackage{fontawesome5} 
\usepackage{dsfont}
\usepackage{titling}
\usepackage{booktabs}
\usepackage{subcaption}
\usepackage[normalem]{ulem}

\newtheorem{corollary}{Corollary}

\newtheorem{proposition}{Proposition}

\newtheorem{axiom}{Axiom}

\newtheorem{theorem}{Theorem}

\theoremstyle{definition}
\newtheorem{example}{Example}
\newtheorem{remark}{Remark}
\newtheorem{definition}{Definition}

\newtheorem{alemma}[]{Lemma}
\newtheorem{aproposition}[]{Proposition}
\DeclareMathOperator{\ind}{ind}

\DeclareMathOperator{\conv}{conv}

\DeclareMathOperator{\D}{D}
\DeclareMathOperator{\supp}{supp}
\DeclareMathOperator{\ext}{ext}

\DeclareMathOperator*{\argmin}{arg\,min}
\DeclareMathOperator{\ex}{ext}

\DeclareMathOperator{\DP}{DP}
\DeclareMathOperator{\Cop}{Cop}

\DeclareMathOperator{\vertices}{vert}

\newcommand{\pind}{p_{\ind}}
\newcommand{\R}{\mathbb{R}}

\newcommand{\res}{\,\rule[-4pt]{0.4pt}{11pt}\,{}} 
\newcommand{\norm}[1]{\left\lVert #1 \right\rVert}
\renewcommand{\epsilon}{\varepsilon}
\newcommand{\Divergence}[3]{\D_{#1}\left( #2  \parallel #3  \right)}
\newcommand{\one}{\mathds{1}}
\newcommand{\Fdn}{\mathcal{F}^{\textup{dn}}}
\newcommand{\Fudn}{\mathcal{F}_u^{\textup{dn}}}
\newcommand{\FvNMdn}{\Fdn}
\newcommand{\md}{\, \mathrm{d}}
\newcommand{\Cind}{C_{\ind}}
\newcommand{\FudnO}{\mathcal{F}_u^{\textup{dn},\circ}}

\title{Dependence uncertainty:\\a decision-theoretic approach}
\author{Gerrit Bauch\thanks{Center for Mathematical Economics, Bielefeld University, PO Box 10 01 31, 33501 Bielefeld, Germany. Email: \href{mailto:gerrit.bauch@uni-bielefeld.de}{gerrit.bauch@uni-bielefeld.de}. Financial support by the German Research Foundation (DFG) [RTG 2865/1 – 492988838].} \  and Lorenz Hartmann\thanks{Basel University, Peter Merian-Weg 6, 4052 Basel, Switzerland. Email: \href{mailto:lorenz.hartmann@unibas.ch}{lorenz.hartmann@unibas.ch}\\%
\faIcon{creative-commons}\faIcon{creative-commons-by}\faIcon{creative-commons-nc}\faIcon{creative-commons-sa} This work first circulated under the title ``Correlation uncertainty:\ a decision-theoretic approach''.%
}}
\date{\today}

\usepackage[round]{natbib}
\begin{document}

\setlength{\droptitle}{-10em}
\maketitle

\begin{abstract}
\noindent%
Individual risks are often better understood than their dependence. We axiomatize choice under dependence uncertainty on finite product state spaces. Preferences reveal utility and marginal beliefs, while acts are evaluated by an act-dependent weighting of worst- and best-case expected utility over all joint distributions consistent with the marginals. We characterize independence-based behavior, which may reflect either justified beliefs or correlation neglect, and introduce dependence premia as well as a comparative ordering of dependence aversion. We classify acts by their exposure to dependence and characterize independence axioms under which the worst–best weight is constant within each class. Finally, we characterize the geometry and extreme points of the dependence uncertainty set, connecting the behavioral analysis to a copula perspective that does not require real-valued random variables.
\smallskip
   
    \noindent%
    \emph{keywords:} dependence uncertainty, product spaces, correlation neglect 
\end{abstract}

\section{Introduction}
Many economic decisions involve several sources of risk with known or subjectively assessed marginal distributions, while their dependence remains uncertain. In dual sourcing, a firm may be able to estimate the disruption risk associated with each supplier, yet lack reliable information about whether the suppliers’ disruptions tend to occur at the same time. A homeowner considering insurance may correctly assess the individual risks of fire and flooding but not know how these damages co-vary and, consequently, how they affect the distribution of total losses. Likewise, an employee owning employer stock may face uncertainty about the dependence between the firm’s performance and her labor income. Marginal distributions alone generally do not determine the risks in such scenarios. We study choice under such \emph{dependence uncertainty}. 

We model distinct sources of uncertainty by a finite product state space $\Omega=\Omega_1\times\cdots\times\Omega_n$, with each $\Omega_i$ representing one such source. Our axioms identify a unique probability measure $p_i$ on each marginal space from preferences over dependence-neutral acts. These marginals may reflect subjective assessments, objective information, or both. By identifying marginal beliefs while leaving the joint distribution unresolved, we isolate dependence uncertainty from other forms of ambiguity. This structure naturally gives rise to the \emph{dependence uncertainty set}
\begin{equation*}
    \mathcal{P} = \mathcal{P}(\Omega; p_1, \ldots, p_n) := \left\{ p \in \Delta(\Omega)
        \mid p \res_{\Omega_i} = p_i \quad \forall i \in \{1, \ldots, n\} \right\},
\end{equation*}
which consists of all joint probability measures on $\Omega$ consistent with the given marginals. An act $f\colon\Omega\to X$ specifies a lottery over prizes in each state, as in \citet{anscombe1963definition}. Our main representation theorem shows that preferences can be represented by
\begin{equation}
    V(f)
    =\alpha(f)\min_{p\in\mathcal P}\int_\Omega u(f)\md p
    +\bigl(1-\alpha(f)\bigr)
      \max_{p\in\mathcal P}\int_\Omega u(f)\md p,
    \label{equation: DU model intro}
\end{equation}
where $u$ is an affine utility function and $\alpha(f)\in[0,1]$ is an act-dependent dependence-aversion index. Conditional on utility and marginal beliefs, this index determines the position of an act's evaluation between its worst- and best-case expected utilities. The key behavioral restriction is independence on \emph{dependence-neutral acts}, i.e., acts whose expected utility is invariant to marginal-preserving changes in the joint distribution for every affine utility. These acts can be evaluated from marginal beliefs alone and serve as benchmarks for acts that are sensitive toward dependence. A monotonicity axiom requires preferences to respect statewise dominance whenever one of the acts is dependence neutral. Together with the remaining axioms, it bounds each act's evaluation by its minimum and maximum expected utility over $\mathcal{P}$. The representation thus separates utility over consequences, beliefs about individual sources of uncertainty, and attitudes toward unresolved dependence.

The dependence uncertainty environment can be combined with several seminal models of ambiguity by restricting the relevant probabilistic assessments to the dependence uncertainty set $\mathcal{P}$. First, an agent who considers a subset $\mathcal{C} \subseteq \mathcal{P}$ of feasible joint distributions and evaluates acts according to their worst-case expected utility obtains a dependence-uncertainty version of Maxmin Expected Utility (MEU) preferences \citep{gilboa1989maxmin}. Second, an agent who holds a second-order probability measure $\mu \in \Delta(\mathcal{P})$ over alternative joint distributions can be modeled within the smooth ambiguity framework of \citet{klibanoff2005smooth}. Third, variational preferences \citep{maccheroni2006ambiguity} fit the framework when their minimal cost function is infinite outside $\mathcal P$. Relative-entropy costs restricted to $\mathcal P$ yield a fixed-marginal counterpart of multiplier preferences
\citep{hansen2001robust}.

One way to resolve dependence uncertainty is to evaluate acts under the independent product of the marginal priors. This benchmark replaces the set of joint distributions consistent with the marginals by the single distribution under which the sources of unknown are independent. When a decision maker treats these unknowns as independent despite relevant evidence of dependence, such behavior is commonly referred to as \emph{correlation neglect}. Correlation neglect has been documented in financial decision making \citep{eyster2010correlation,eyster2016correlation,kallir2009neglect} and in belief formation \citep{enke2019correlation}. We use the more neutral term \emph{independence-based preferences}, since evaluation under the independent
product need not reflect neglect:\ independence may instead be objectively justified or serve as a deliberate probabilistic benchmark. Within the intersection of MEU and dependence-uncertainty preferences, we provide a behavioral separation axiom that characterizes when the relevant set of priors reduces to the independent product of the marginal priors. In this case, preferences reduce to subjective expected utility under independence.

We also develop two complementary measures of attitudes toward dependence. The first, the \emph{dependence premium}, is the monetary compensation associated with replacing the decision maker's evaluation of an act under dependence uncertainty by its expected-utility evaluation under the independent product of the marginal priors. The second is a behavioral comparison of \emph{dependence aversion}, based on \cite{yaari1969some}: holding risk preferences and marginal beliefs fixed, a decision maker is more dependence averse than another if she is less willing to choose a dependence-sensitive act over a dependence-neutral benchmark. Within our dependence-uncertainty model, these two notions induce the same ordering of preferences. Non-negative dependence premia for all acts are equivalent to \emph{absolute dependence aversion}. We characterize how these notions specialize within the Maxmin Expected Utility, smooth ambiguity, and variational models.

Moreover, we study additional independence requirements that further restrict the act-dependent ambiguity index. We distinguish acts that have the same cardinal exposure to dependence from acts whose dependence exposure differs only by a positive rescaling. The former acts differ only through components that are themselves dependence neutral, whereas the latter preserve the ordinal ranking of dependence exposure. We characterize both relations and show that strengthening restricted independence requires the ambiguity index $\alpha(\cdot)$ to be constant within the corresponding classes of acts. These axioms are dependence-uncertainty counterparts of Weak Certainty Independence \citep{maccheroni2006ambiguity} and Certainty Independence \citep{gilboa1989maxmin}, respectively: rather than using constant acts as the relevant benchmarks, they use dependence-neutral acts. The results therefore clarify when attitudes toward dependence uncertainty may vary across acts and when they must instead be invariant to changes that do not alter an act's dependence exposure.

Finally, we mathematically characterize the dependence uncertainty set $\mathcal{P}$ in two ways. First, as the intersection of an affine space, generated by dependence structures and the simplex. This yields an explicit formula on the dimension of $\mathcal{P}$ and thus of the scope of dependence given the state space. Second, We characterize the extreme points of $\mathcal{P}$ as measures with minimal support and local maximizers of a class of divergences that measure the distance to the independent product. We conclude by deriving how the set of dependence structures generalizes the statistical object of a copula to un-ordered state spaces.

\paragraph{Related literature.}

Our analysis relates to the distinction between ambiguity about individual sources of risk and ambiguity about their relationship. \citet{epstein2019ambiguous} provide experimental evidence that uncertainty about the joint distribution of multiple random events is behaviorally relevant in addition to classical Ellsbergian ambiguity about their individual distributions. We provide a general decision-theoretic treatment of this latter, dependence-related component of ambiguity: marginal beliefs are identified, while the joint distribution remains unresolved. %
Relatedly, \citet{ellis2025correlation} studies decision makers who are concerned that they may misunderstand the joint realizations of actions or assets. We instead take a finite product state space as primitive, identify marginal beliefs from preferences over acts, and allow act-dependent responses to unresolved dependence.

Our work is related to \citet{ghirardato2004differentiating} along two dimensions. First, both representations combine worst- and best-case expected utility through an act-dependent aversion index. In their model, the bounds are taken over the Bewley set representing the decision maker's derived unambiguous preference relation \citep{bewley2002knightian}. Our bounds range over all joint probabilities consistent with the revealed marginal beliefs, thereby isolating dependence as the source of unresolved uncertainty. %
Accordingly, our index measures aversion to dependence uncertainty rather than ambiguity in general. Second, our dependence-invariance refinements parallel their requirement that the ambiguity-aversion index be constant within equivalence classes of acts. Whereas their classes capture exposure to the decision maker's revealed ambiguity, ours are preference-independent and distinguish cardinal from ordinal exposure to dependence. We characterize when the dependence-aversion index is constant within each class through counterparts of Weak Certainty Independence and Certainty Independence in which dependence-neutral acts replace constant acts.

Our characterization of independence-based preferences relates to the axiomatic literature on subjective stochastic independence. \cite{grabisch2023subjective} derive SEU with a product probability measure in a Savage framework. Their Subjective Independence axiom implies, under suitable richness conditions, that conditioning on one source does not alter rankings between acts depending only on the other source, an invariance property closely related to our Subspace Separation. \cite{ceron2021stochastic} study several notions of stochastic independence for Bewley and maxmin preferences in an Anscombe--Aumann framework. The Stochastic Independence axiom imposes an ex‑ante aggregation requirement that must hold for all arbitrary acts, whereas our Subspace Separation axiom is milder: it only requires that conditioning acts that live on one marginal space on an event of the residual space preserves the ranking between acts that differ within a single subspace.

Independence-based preferences are also related to correlation neglect, whereby decision makers evaluate jointly determined outcomes as if their underlying sources were independent. Experimental evidence documents such behavior in portfolio allocation \citep{kallir2009neglect,eyster2016correlation} and belief updating from correlated information sources \citep{enke2019correlation}. Correlation neglect also arises in strategic settings. In cursed equilibrium, players correctly anticipate opponents' action frequencies but underweight their dependence on private
information \citep{eyster2005cursed}. \citet{preker2025cursedequilibriaknightianuncertainty} combines cursedness with ambiguity about type distributions. In his trading game, ambiguity can increase trade among cursed, uncertainty-averse players.

Our comparative analysis adapts the benchmark-comparison approach of \citet{ghirardato2002ambiguity}, replacing constant acts with dependence-neutral benchmarks. It compares decision makers by their willingness to accept dependence-sensitive acts relative to these benchmarks. This differs from the correlation aversion studied by \citet{epstein1980increasing}, which concerns expected-utility rankings of increases in positive dependence. 

More recently, \citet{stanca2026recursive} studies correlation aversion under recursive preferences, highlighting the trade-off between hedging and early resolution of uncertainty. His persistence premium measures the consumption share sacrificed to eliminate persistence. Our dependence premium shares the independence benchmark but measures monetary compensation in a static setting and allows for ambiguous dependence.

The fixed-marginal, uncertain-dependence structure also appears in several economic applications. \citet{carroll2017robustness} studies a screening problem in which the principal knows marginal type distributions and evaluates mechanisms by worst-case profit over all compatible joint distributions. In portfolio choice, \citet{jiang2022portfolio} consider investors who know assets' marginal return characteristics but are ambiguous about their correlation matrix, while \citet{levy2022combining} study forecast aggregation under ambiguity about dependence across information sources. %
Coarse memory provides another source of missing dependence information: in \citet{bilotta2024coarse}, voters remember policy and outcome frequencies but forget their association, allowing politicians to exploit this informational gap through plausible
narratives.

Finally, our fixed-marginal uncertainty set connects to robust optimization and risk aggregation. \citet{agrawal2012price} bound the performance loss from assuming independence in stochastic optimization, while \citet{bernard2014risk} derive bounds on aggregate risks with prescribed marginals and unspecified dependence. \citet{puccetti2015extremal} survey extremal dependence and its role in probabilistic optimization. Our focus is on the preference foundations and geometry of the underlying dependence uncertainty set.

\bigskip
The remainder of the article is organized as follows. Section \ref{Section: Preliminaries and Notation} briefly introduces the mathematical setup and notations. In Section \ref{Section: P and Q}, we describe the dependence uncertainty set, formally introduce dependence structures, and provide an additive decomposition of every probability into its independent product and a dependence structure. %
Our main contribution, the definition and axiomatic characterization of dependence uncertainty, is developed in Section \ref{section: model}. %
We apply this model to maxmin expected utility, smooth ambiguity, and variational preferences and provide an axiomatization of independence-based preferences. %
Section \ref{section: comparative dependence uncertainty and dependence premia} introduces the concept of dependence premium and comparative dependence aversion. 
Section \ref{Section: dependence-invariance} introduces cardinal and ordinal notions of dependence equivalence and provides behavioral foundations for treating acts with the same exposure to dependence alike. In Section \ref{Section: Math properties of P}, we characterize the dependence uncertainty set and its extreme points, and show how our concept of dependence structures generalizes the statistical concept of copulae.
\section{Preliminaries and notation}\label{Section: Preliminaries and Notation}
Let $\Omega$ be a finite \emph{state space} that is the product of \emph{subspaces} $\Omega_1, \dots, \Omega_n$ such that $\Omega = \Omega_1 \times \dots \times \Omega_n$. A \emph{state} is of the form $\omega=(\omega_1, \dots, \omega_n)$, where $\omega_i\in \Omega_i$. A set of states is called an \emph{event} and denoted by $E \subseteq \Omega$, likewise $E_i \subseteq \Omega_i$ for subspaces. $\Omega_{-i}:=\Omega_1 \times \dots \times \Omega_{i-1} \times \Omega_{i+1} \times \dots \times \Omega_n$ is the Cartesian product of all subspaces but $\Omega_i$. %
For $\omega_i \in \Omega_i$  we define $[\omega_i] := \{\omega_i\} \times \Omega_{-i}\subseteq \Omega$, i.e., $[\omega_i]$ coincides with the set of states that agree with $\omega_i$ in the $i$th index. Analogously, $[E_i], [\omega_{-i}], [E_{-i}] \subseteq \Omega$ are defined for $E_i \subseteq \Omega_i, \omega_{-i} \in \Omega_{-i}, E_{-i} \subseteq \Omega_{-i}$.

Denote by $\Delta(\Omega)$ the set of all probability measures $p$ on $\Omega$. For a probability $p \in \Delta(\Omega)$ and $i \in \{1, \ldots, n\}$, the \emph{marginal} $p \res_{\Omega_i} \in \Delta(\Omega_i)$ of $p$ on $\Omega_i$ is given by %
$
    p \res_{\Omega_i}(\omega_i) := p([\omega_i]) = \sum_{\omega_{-i} \in \Omega_{-i}} p(\omega_i,\omega_{-i}),
$
for any $\omega_i \in \Omega_i$. The support $\supp(p_i)$ is the set of states in $\Omega_i$ for which $p_i$ is non-zero. A \emph{signed measure} is a ($\sigma$-)additive mapping $q \colon 2^\Omega \to \R$, i.e., $q(E \dot{\cup}F) = q(E) + q(F)$ for all disjoint $E,F \subseteq \Omega$. By considering the evaluations state-wise, probabilities and signed measures can be treated as vectors in $\R^{\# \Omega}$. For real-valued functions, the integrals over (signed) measures are defined as usual, coinciding with the scalar product in the Euclidean space. %

We consider the framework of \cite{anscombe1963definition}. We fix a set $Z$ of prizes and let $X$ denote consequences, which are finite-valued lotteries over $Z$. We consider \emph{acts}, i.e., maps $f \colon \Omega \to X$, the set of which is denoted by $\mathcal{F}$. An act $f\in \mathcal{F}$ is called \emph{independent of $\Omega_{-i}$} if $f$ is constant on $[\omega_i]$ for all $\omega_i \in \Omega_i$. The set of $\Omega_{-i}$-independent acts is denoted by $\mathcal{F}_i$ and identifies with the set $\{f_i:\Omega_i \rightarrow X\}$ via the natural embedding. For two acts $f,g\in \mathcal{F}$ and an event $E\subseteq \Omega$, $f_Eg$ denotes the act that results in $f(\omega)$ on $E$ and $g(\omega)$ on $E^c \colon= \Omega\setminus E$. %
Preferences over all acts are indicated by a binary relation $\succcurlyeq$ on $\mathcal{F}$. An event $E \subseteq \Omega$ is called \emph{null}, if $f_Eg \sim g$ for all $f,g \in \mathcal{F}$, otherwise we say that $E$ is \emph{non-null}. In representations of a preference relation, \emph{utility functions} $u \colon X \to \R$ map outcomes to utilities.
\section{Dependence Uncertainty Set and Dependence Structures}
\label{Section: P and Q}

This section introduces two crucial objects used throughout the paper. The \emph{dependence uncertainty set} collects all joint distributions consistent with given marginals. \emph{Dependence structures} describe how a probability measure on a product space differs from the independent product of its marginals. We motivate these objects by means of a simple example and then introduce them formally. We subsequently show that every joint distribution admits a unique decomposition into its independent product and a dependence structure.

\begin{example}[Procurement]\label{Example: dual sourcing}
A producer orders inputs $x_i\geq 0$ from suppliers $i$, $i\in \{1,2\}$. At the delivery date, supplier $i$ is either up ($U_i$) and fills its order, or disrupted ($D_i$) and cannot deliver. Past supplier-specific performance records the probabilities of $U_1$ and $U_2$. However, the dependence of $U_1$ and $U_2$ is not fully recorded, the producer thus faces \emph{dependence uncertainty}.\footnote{Suppliers may face distinct local or logistical risks---for example, relying on road transport or on sea transport---but also common shocks, such as upstream shortages or trade restrictions. Incomplete panel data may capture each supplier's marginal operational reliability yet provide too few matched observations to reliably identify dependence between their operational states. } Table \ref{Table: dual sourcing distributions} parametrizes all joint distributions consistent with probability $3/4$ of $U_1$ and probability $2/3$ of $U_2$. The parameter $a$ shifts probability mass between the two diagonals while leaving marginals unchanged. Independence corresponds to $a=0$. A positive $a$ makes the suppliers more likely to be up  together as well as more likely to be disrupted together. Dependence uncertainty is reflected by the parameter $a$ being unknown.

\begin{table}[htb]
    \centering
    \begin{tabular}{c||c|c}
         $p$ & $U_2$ ($2/3$) & $D_2$ ($1/3$) \\\hline\hline
         $U_1$ ($3/4$) & $\tfrac{1}{2}+a$ & $\tfrac{1}{4}-a$ \\\hline
         $D_1$ ($1/4$) & $\tfrac{1}{6}-a$ & $\tfrac{1}{12}+a$
    \end{tabular}
    \caption{All joint distributions on $\Omega = \{U_1,D_1\}\times\{U_2,D_2\}$ with marginals $p_1(U_1)=3/4$ and $p_2(U_2)=2/3$, parametrized by $a\in[-1/12,1/6]$.}
    \label{Table: dual sourcing distributions}
\end{table}
\end{example}

Let $\Omega=\Omega_1\times\cdots\times\Omega_n$ be a finite product state space. For any profile of marginal probabilities $p_i \in \Delta(\Omega_i)$, define the associated dependence uncertainty set as the set contains all joint distributions consistent with these marginals.\footnote{In the representation theorem below, these marginal probabilities are not primitive: they are uniquely identified from preferences over dependence-neutral acts. } 

\begin{definition}[Dependence Uncertainty Set]
    The \emph{dependence uncertainty set (given $p_1,  \ldots, p_n$)} is
\begin{equation}
    \mathcal{P}(\Omega; p_1, \ldots, p_n) := \left\{ p \in \Delta(\Omega) \mid p \res_{\Omega_i} = p_i \, \forall i \in \{1, \ldots, n\}\right\}.\label{eq: general definition P}
\end{equation}
\end{definition}

If the marginals and the state space are clear from the context, we simply write $\mathcal{P}$. The dependence uncertainty set always contains the \emph{independent product} $\pind := p_1 \otimes \ldots \otimes p_n$, defined by $\pind(\omega) = \prod_{i=1}^n p_i(\omega_i)$, where $\omega=(\omega_1, \ldots, \omega_n)$. If the $p_i$ have full support, or if restricting the state space to that, $\pind$ is a relative-interior point of $\mathcal{P}$. %
Table \ref{Table: dual sourcing distributions} describes the set $\mathcal{P}$ within Example \ref{Example: dual sourcing} and we obtain $\pind = (\begin{smallmatrix}
    1/2 & 1/4\\
    1/6 & 1/12
\end{smallmatrix})$ for $a=0$.\medskip

Different elements of $\mathcal{P}$ share the same marginal information but encode different dependence among the marginal spaces. To isolate this dependence, consider the difference between a joint distribution and its independent product. Such a difference is a signed measure whose marginal mass is zero.

\begin{definition}[Dependence Structure]\label{Definition: dependence structure}
    A signed measure $q \colon 2^\Omega \to \R$ with $q([\omega_i]) = 0$ for all $i \in \{1, \ldots, n\}$ and $\omega_i \in \Omega_i$ is called a \emph{dependence structure}. The set of all dependence structures is denoted by $Q = Q(\Omega)$.
\end{definition}
Dependence structures formally capture how the mass of a probability measure can be re-allocated without changing the marginal distributions. %
In Example \ref{Example: dual sourcing}, we can write any $p = \pind + q$ where the signed measure $q = (\begin{smallmatrix}
   a & -a\\
    -a &a
\end{smallmatrix})$
is a dependence structure described by the single parameter $a \in \R$.\footnote{Indeed, these \emph{rectangular operations} serve as general prototype for elementary dependence structures, that turn out to generate all dependence structures, cf.\ Section \ref{Section: Math properties of P}.} %
We show that indeed any probability can be uniquely decomposed into its independent product and a dependence structure like in Example \ref{Example: dual sourcing}.

\begin{proposition}[Dependence Decomposition]\label{Proposition: decomposition}
    Every probability $p \in \Delta(\Omega)$ with marginals $p_1, \ldots, p_n$ has a unique additive decomposition $p = p_1 \otimes \ldots \otimes p_n + q$ with a dependence structure $q \in Q$. Vice versa, for any $q \in Q$, such that $p := p_1 \otimes \ldots \otimes p_n + q \in \Delta(\Omega)$, $p$ has marginals $p_1, \ldots, p_n$. In particular,
    \begin{equation*}
        \mathcal{P} = (\pind+Q)\cap\Delta(\Omega).
    \end{equation*}
\end{proposition}
\begin{proof}
    All proofs are relegated to the appendix.
\end{proof}

Once the marginals are fixed, uncertainty about the joint distribution thus corresponds precisely to uncertainty about the feasible dependence structure. Section \ref{Section: Math properties of P} develops the geometry of these objects. It constructs a spanning system for $Q$, calculates its dimension, and characterizes the extreme points of the polytope $\mathcal P$ as measures with minimal support and locally maximal divergence.

\section{A choice-theoretic investigation of dependence uncertainty}\label{section: model}

This section develops a behavioral foundation of preferences in face of dependence uncertainty. We now derive the marginal beliefs and the associated dependence uncertainty set from preferences. The representation will identify marginal priors $p_i$, after which $\mathcal P=\mathcal P(\Omega;p_1,\ldots,p_n)$ captures all unresolved uncertainty about dependence. Our main model evaluates an act $f$ according to 
\begin{equation}
     V(f)=\alpha(f)\min_{p\in\mathcal P}\int_\Omega u(f)\md p
      +(1-\alpha(f))\max_{p\in\mathcal P}\int_\Omega u(f)\md p. 
\end{equation}
Here, $u$ reflects the decision maker's risk preference, while $\alpha(f)$ describes how she responds to uncertainty about dependence. We first identify acts whose evaluation does not hinge on dependence and then axiomatize the above model. We subsequently locate established ambiguity models within our theory and characterize behavior based on the independent product of the marginal beliefs.

\subsection{Dependence-neutral acts}
\label{sec:dependence-neutral acts}

An act is \emph{dependence neutral} if its expected utility is unaffected by a change in the dependence structure. Put differently, the only probabilistic information needed for its evaluation are the marginal distributions of the subspaces themselves. Such acts serve as the anchoring objects in our axioms.

\begin{definition}[Dependence neutrality of acts]
\label{definition:dependence-neutral-acts}
Let $f\in\mathcal F$.
\begin{enumerate}[(i)]
    \item The act $f$ is \emph{dependence neutral given
    $u\colon X\to\mathbb R$} if
    \begin{equation*}
        \int_\Omega u(f)\,\mathrm dq=0
        \qquad\text{for every }q\in Q.
    \end{equation*}
    The set of such acts is denoted by $\Fudn(\Omega)$. If the state space is clear from the context, we simply write $\mathcal F_u^{\mathrm{dn}}$.\label{item: definition dependence-neutral acts}

    \item The act $f$ is \emph{dependence neutral} if it is dependence neutral given every affine $u\colon X\to\mathbb R$. The set of such acts is denoted by $\Fdn = \Fdn(\Omega)$.
\end{enumerate}
\end{definition}

Dependence neutrality given $u$ is equivalent to requiring $\int u(f)\,\mathrm dp=\int u(f)\,\mathrm dp'$ whenever measures $p$ and $p'$ have the same marginals, as can be seen by taking the difference of these expected utilities.

The following result gives a utility-free characterization of the set $\FvNMdn$. Recall that $\mathcal F_i$ denotes the set of acts that depend only on marginal space $\Omega_i$.

\begin{proposition}\label{Proposition: characterization dependence neutrality for all notions}
An act $f\in\mathcal F$ is dependence neutral if and only if there exist $f_i\in\mathcal F_i$ and $\gamma_i\geq0$, $i=1,\ldots,n$, with $\sum_{i=1}^n\gamma_i=1$ such that
\begin{equation*}
    f=\sum_{i=1}^n\gamma_i f_i.    
\end{equation*}
\end{proposition}

In presence of marginal distributions $p_1, \ldots, p_n$, dependence neutrality given $u$ is not necessary to ensure that the act's evaluation is constant over $\mathcal{P}$ if the marginals do not have full support. To establish a one-to-one correspondence to acts that are constant on $\mathcal{P}$, we suitably enlarge the set $\Fudn$ as follows.
\begin{definition}
    Let $p_1, \ldots, p_n$ and $u \colon X \to \R$ be given. Denote by $\Omega^\circ = \Omega^\circ(p_1, \ldots, p_n) = \Omega \setminus \pind^{-1}(0)$ the restriction of $\Omega$ to the support of $\pind$. We denote by
    \begin{equation*}
        \FudnO = \FudnO(\Omega; p_1, \ldots, p_n) := \left\{ f \in \mathcal{F} \mid f \res_{\Omega^\circ} \in \Fudn(\Omega^\circ(p_1, \ldots, p_n))\right\}
    \end{equation*}
    the set of acts that are dependence neutral given $u$ when restricted to $\Omega^\circ$.
\end{definition}
\begin{proposition}\label{Proposition: Fudn adjusted to marginals}
    We have $\Fudn(\Omega) \subseteq \FudnO(\Omega;p_1,\ldots,p_n)$. Furthermore, an act $f \in \mathcal{F}$ is in $\FudnO(\Omega; p_1, \ldots, p_n)$ if and only if $\min_{p \in \mathcal{P}(\Omega; p_1, \ldots, p_n)} \int_{\Omega} u(f) \md p = \max_{p \in \mathcal{P}(\Omega; p_1, \ldots, p_n)} \int_{\Omega} u(f) \md p$, i.e., $\FudnO(\Omega;p_1, \ldots, p_n)$ contains exactly those acts whose expected utility under $u$ is invariant over $\mathcal{P}(\Omega; p_1, \ldots, p_n)$
\end{proposition}

Table \ref{Table:example dependence-neutral acts} illustrates the concept of dependence neutrality.

\begin{table}[htb!]
    \centering
    \begin{tabular}{c||c|c}
        $f$ & $U_2$ & $D_2$ \\ \hline\hline
        $U_1$ & $x_1$ & $x_1$ \\ \hline
        $D_1$ & $x_0$ & $x_0$
    \end{tabular}
    \qquad
    \begin{tabular}{c||c|c}
        $g$ & $U_2$ & $D_2$ \\ \hline\hline
        $U_1$ & $x_1$ & $x_{1/2}$ \\ \hline
        $D_1$ & $x_{1/2}$ & $x_0$
    \end{tabular}
    \qquad
    \begin{tabular}{c||c|c}
        $h$ & $U_2$ & $D_2$ \\ \hline\hline
        $U_1$ & $x$ & $y$ \\ \hline
        $D_1$ & $y$ & $y$
    \end{tabular}
    \caption{Consider the procurement example from Section \ref{Section: P and Q}. Assume for $f$ and $g$ that deliveries are perfect substitutes, i.e., the firm cares only about the total quantity received. Consequences are lotteries over delivered quantities, with $x_z$ denoting the degenerate lottery delivering $z \in \R_{\geq 0}$ units. The firm can place its entire order with supplier 1 (sole sourcing, act $f$) or order half a unit from each supplier (split sourcing, act $g$). Sole sourcing depends only on supplier 1's availability and is therefore dependence neutral for every utility function, i.e., $f\in\mathcal \Fdn$. Split sourcing is dependence neutral given $u$ if and only if $u(x_1)+u(x_0)=2u(x_{1/2})$. This condition holds when the firm evaluates lotteries by expected delivered quantity, but it fails when the value of delivered quantity, ($z \mapsto u(x_z)$), is, e.g., strictly concave. Thus, split sourcing belongs to $\Fudn$ for some affine utility functions, but not to $\mathcal \Fdn$. The act $h$ reflects split sourcing when the inputs are perfect complements, i.e., both suppliers need to be up to produce the good (outcome $x$), otherwise there is nothing produced (outcome $y$). The act $h$ is only dependence neutral given a utility function with $u(x) = u(y)$. Further, it is an element of $\Fdn$ whenever $x=y$. If in addition there are marginals with, say, $p_1(U_1) = 1$ and $p_2(U_2) = \tfrac{1}{2}$, i.e., $p_1$ does not have full support, the set $\mathcal{P}$ is a singleton. The expected utility of all acts is trivially constant over $\mathcal{P}$, thus $\FudnO(\Omega; p_1, p_2)=\mathcal{F}(\Omega)$ for all $u$.}
    \label{Table:example dependence-neutral acts}
\end{table}

\subsection{Representation}\label{sec:axioms and representation}

We impose the following set of axioms on a binary relation $\succcurlyeq$ on $\mathcal{F}$.

\begin{axiom}[Non-trivial Weak Order]\label{axiom: non-trivial weak order}
    $\succcurlyeq$ is complete, transitive and non-trivial.
\end{axiom}

\begin{axiom}[Unboundedness]\label{axiom: unboundedness} There exist $x_1, x_2\in X$ with $x_1\succ x_2$ such that, for all $\lambda \in (0,1)$, there exist $y_1, y_2\in X$ such that $x_2 \succ \lambda y_1 + (1 - \lambda)x_1$ and $\lambda y_2 + (1 -\lambda)x_2 \succ x_1$.
\end{axiom}

\begin{axiom}[$\FvNMdn$-Bracketed Archimedean]\label{axiom: dn archimedean}
   For all $g\in \mathcal{F}$ and all $f,h\in \FvNMdn$, if $f\succ g \succ h$, then there exist $\lambda, \kappa \in (0,1)$ such that 
    \begin{equation*}
        \lambda f + (1-\lambda) h \succ g \succ \kappa f + (1-\kappa) h. 
    \end{equation*}
\end{axiom}

\begin{axiom}[$\FvNMdn$ - Independence]\label{axiom: dn independence}
    For all $f, g, h \in \FvNMdn$, $\lambda\in (0,1)$,
    \begin{equation*}
        f\succcurlyeq g \iff \lambda f +(1-\lambda) h \succcurlyeq \lambda g +(1-\lambda) h.
    \end{equation*}
\end{axiom}

\begin{axiom}[$\FvNMdn$-Bracketed Dominance]\label{axiom: dn dominance}
    For all $f\in \mathcal{F}$, $g\in \FvNMdn$, 
\begin{equation*}
    \big[ \, (\forall \omega \ f(\omega)\succcurlyeq g(\omega) ) \implies f\succcurlyeq g \, \big] \qquad \text{and} \qquad \big[ \, (\forall \omega \ f(\omega)\preccurlyeq g(\omega) )  \implies f\preccurlyeq g \, \big].
\end{equation*}
\end{axiom}
Axioms \ref{axiom: non-trivial weak order} is standard. Axiom \ref{axiom: unboundedness} requires sufficiently extreme outcomes to overturn a fixed strict ranking at any positive mixture weight. Axiom \ref{axiom: dn archimedean} requires the usual Archimedean condition whenever acts are wedged by dependence-neutral acts. Axiom \ref{axiom: dn independence} requires the classic independence condition on $\FvNMdn$. Axiom \ref{axiom: dn dominance} asserts monotonicity whenever one of the two considered acts is in $\FvNMdn$. 

\begin{theorem}[Dependence Uncertainty]\label{theorem: dependence uncertainty}
    Let $\succcurlyeq$ be a preference relation on $\mathcal{F}$. The following are equivalent:
    \begin{enumerate}
        \item $\succcurlyeq$ satisfies Axioms \ref{axiom: non-trivial weak order} - \ref{axiom: dn dominance}. 
        \label{item: representation result axioms} 
        \item $\succcurlyeq$ admits a dependence uncertainty representation, i.e., there exists an unbounded affine $u\colon X\to \mathbb{R}$, an ambiguity index $\alpha\colon\mathcal{F} \to [0,1]$ and $p_1, \ldots, p_n \in \Delta(\Omega_1), \ldots, \Delta(\Omega_n)$ such that $\succcurlyeq$ is represented by the functional 
    \begin{equation*}
        V(f) = \alpha(f) \min\limits_{p\in \mathcal{P}} \int_\Omega u(f) \md p + (1-\alpha(f)) \max\limits_{p\in \mathcal{P}} \int_\Omega u(f) \md p, 
    \end{equation*}   
    where $\mathcal{P}:=\mathcal{P}(\Omega;p_1, \ldots, p_n)$.\label{item: representation result representation}
     \end{enumerate}   
    Furthermore, $u$ is unique up to positive affine transformations, $p_1, \ldots, p_n$ are unique and $\alpha$ is unique on $\mathcal{F} \setminus \FudnO(\Omega; p_1, \ldots, p_n)$.
\end{theorem}

Throughout, we normalize without loss $\alpha(f)=1/2$ for every $f\in \FudnO$, making the index absolutely unique. 

The logic of the representation is as follows. Axioms \ref{axiom: non-trivial weak order}, \ref{axiom: dn archimedean} and \ref{axiom: dn independence} imply, through the Anscombe-Aumann theorem, a subjective expected utility representation on each $\mathcal{F}_i$, thus identifying an affine utility function $u(X) = \R$ and marginal beliefs $p_i$ within each marginal space. This eliminates all probabilistic uncertainty ``within subspaces". Through Proposition \ref{Proposition: characterization dependence neutrality for all notions} this evaluation is extended to all dependence-neutral acts. Finally, Axiom \ref{axiom: dn dominance} ensures that every act is between its worst and best expected utility over the dependence-uncertainty set $\mathcal{P}$.\footnote{Indeed, this is only achieved in presence of Axiom \ref{axiom: unboundedness}. As a counterexample, assume $u(X) = [0,1]$, $\Omega = \{ \omega_1, \omega_2, \omega_3 \}^2$ with uniform marginals and define a preference relation by $V(f) = \max\{ \int_{\Omega} u(h) \md \pind \mid u(h) \leq u(f), h \in \FvNMdn\}$, which fulfills Axioms \ref{axiom: non-trivial weak order} and \ref{axiom: dn archimedean}\textemdash \ref{axiom: dn dominance}. Then, an act $f$ with utility profile $u(f) = \left(\begin{smallmatrix} .2 & .2 & .1\\ 1 & .7 & .4\\ 0 & 0 & 1 \end{smallmatrix}\right)$ yields $V(f) = \tfrac{1}{6} < \tfrac{1}{5} = \min \{ \int_{\Omega} u(f) \md p \mid p \in \mathcal{P} \}$ and thus does not fall in the bounds of the dependence uncertainty representation \eqref{equation: DU model intro}.} The weight $\alpha(f)$ records its position between these bounds.

\subsection{Dependence uncertainty in the maxmin expected utility, smooth ambiguity and variational model}\label{section: MEU and smooth}

We characterize the restrictions that dependence uncertainty imposes on three well-established models of ambiguity preferences. Our proposition shows that the resulting probabilistic uncertainty must be entirely attributable to dependence. Intuitively, within each of the ambiguity models considered, dependence uncertainty requires preferences to be ``concentrated on $\mathcal{P}$''. 

In the \emph{maxmin expected utility} (MEU) model of \cite{gilboa1989maxmin}, acts are evaluated according to the functional
\begin{equation*}
    V(f)= \min_{p\in \mathcal{C}} \int_\Omega u(f) \md p, 
\end{equation*}
where $u\colon X\to \mathbb{R}$ is a non-constant affine function and $\mathcal{C}\subseteq \Delta(\Omega)$ is a non-empty, convex, compact set of priors. The set $\mathcal{C}$ is frequently interpreted as the set of probability distributions that the DM cannot rule out.

In the \emph{smooth ambiguity} model of \cite{klibanoff2005smooth}, acts are evaluated according to the functional
\begin{equation*}
    V(f) = \int_{\Delta(\Omega)} \phi \left(\int_{\Omega} u(f) \md p\right) \md\mu,
\end{equation*}
where $u\colon X\to \mathbb{R}$ is a non-constant affine function, $p \in \Delta(\Omega)$ denotes a probability measure on $\Omega$, and $\phi \colon u(X) \to \mathbb{R}$ is a continuous, strictly increasing function, where $u(X)$ denotes the range of $u$. %
The parameter $\mu$ is a probability distribution over $\Delta(\Omega)$, i.e., a second-order probability distribution. It reflects the ambiguity that the DM perceives about the likelihood of elements of $\Delta(\Omega)$.\medskip

In the \emph{variational model} of \citet{maccheroni2006ambiguity}, acts
are evaluated by
\[
    V(f)=\min_{p\in\Delta(\Omega)}
    \left\{\int_\Omega u(f)\md p+c(p)\right\},
\]
where $u\colon X\to\mathbb R$ is nonconstant and affine, and
$c\colon\Delta(\Omega)\to[0,\infty]$ is grounded, convex, and lower
semicontinuous. Grounded means that $\min_p c(p)=0$. We restrict attention to the unique \emph{minimal cost function}, defined by $c^*(p) = \sup_{f \in \mathcal{F}} \{ V(f) - \int_{\Omega} u(f) \md p \}$, which is unique for a fixed utility normalization (\citet{maccheroni2006ambiguity}, Theorem 3).

\begin{proposition}[MEU, smooth and variational dependence uncertainty]\label{prop: dependence uncertainty MEU and smooth}
Let $\succcurlyeq$ exhibit dependence uncertainty with marginals $p_1, \ldots, p_n$.
\begin{enumerate}[(i)]
    \item If in addition $\succcurlyeq$ is MEU with prior set $\mathcal{C}$, then $\mathcal{C}\subseteq \mathcal{P}(\Omega; p_1, \ldots, p_n)$.\label{item: MEU subset dependence uncertainty}
    \item If in addition $\succcurlyeq$ is smooth with non-affine ambiguity index $\phi$ and second-order prior $\mu$, then $\mu(\mathcal{P}(\Omega; p_1, \ldots, p_n)) = 1$.\footnote{If $\phi$ is affine, the smooth preference is equivalent to an SEU preference w.r.t.\ the compound lottery, which may be a convex combination of priors with marginals unequal to $p_i$.}\label{item: smooth subset dependence uncertainty}
    \item If in addition $\succcurlyeq$ is variational with minimal cost function $c^*$, then $c^*(p)=\infty$ for all $p\notin \mathcal{P}(\Omega;p_1, \ldots, p_n)$ and $c^*(p)< \infty$ for some $p\in \mathcal{P}(\Omega;p_1, \ldots, p_n)$. \label{item: variational characterization dependence uncertainty}
\end{enumerate}
\end{proposition}

\subsection{A characterization of independence-based preferences}\label{Section: dependence neglect}

Basing a decision on the independent product $\pind=p_1\otimes\ldots\otimes p_n$ allows for two distinct behavioral interpretations. Firstly, a decision maker may use it despite evidence of dependence, in which case the behavior reflects \emph{dependence neglect}. Secondly, she may know that the marginal spaces are independent, and acts accordingly. We use the neutral term \emph{independence-based preferences} for such preferences, represented by
\begin{equation*}
    V_{\pind}(f)=\int_\Omega u(f)\md \pind.
\end{equation*}

We characterize independence-based preferences at the intersection of MEU and dependence uncertainty preferences by means of a behavioral axiom.

To motivate the axiom, consider potential fire ($B$) and flood ($F$) threats to a property. Let $f_B$ and $f^\prime_B$ be two indemnity schedules contingent on fire but not on flood. Now, consider the following modified indemnity schedules:\ in absence of a flood, the original schedule applies; in case of a flood, the insurance denies coverage even if the covered fire peril occurred in sequence.\footnote{A policy provision that bars coverage when a loss results from a combination of covered and excluded perils occurring simultaneously or in sequence is known as an \emph{anti-concurrent causation clause} and common in property and casualty insurance contracts. For instance, \emph{Insurance Services Office, Inc., Homeowners 3—Special Form, Form HO 00 03 05 11, § I, Exclusions} states ``We do not insure for [the following types of] loss [\ldots], regardless of any other cause or event contributing concurrently or in any sequence to the loss.''} If $B$ and $F$ are independent, the probability of $B$ is equal to the probability of $B$ conditional on $\neg F$. Thus, this modification should not reverse the ranking. Under dependence, however, conditioning on $\neg F$ may alter the probability of fire and thereby reverse the ranking.\footnote{Such dependence is realistic. Indeed, the IPCC concludes that human-induced climate change has increased the probability of compound extreme events and documents changes in both fire weather and compound flooding in some regions \citep[Chapter~11, especially Section~11.8]{seneviratne2021weather}. Thus, hazards such as fire and flooding, although not directly causally related, may be statistically dependent through common climatic drivers. An insurer may account for such dependence using richer data and catastrophe models, whereas a homeowner may evaluate the hazards separately.}

Our axiom rules out preference reversals for such modifications. Formally, the axiom asserts that the ranking of two $\mathcal{F}_i$-acts is unaltered when replacing the consequences on complements of events $[E_{-i}] = \Omega_i \times E_{-i}$ by a constant, i.e., if we ``condition'' on $[E_{-i}]$. Table \ref{Table: illustration subspace separation} illustrates the axiom by means of the insurance story.

\begin{axiom}[Subspace Separation]\label{axiom:subspace separation}
    For all $f_i,g_i \in \mathcal{F}_i$, $x \in X$, non-null $E_{-i} \subseteq \Omega_{-i}$,
    \begin{equation*}
        f_i\succcurlyeq g_i \iff {f_{i}}_{[E_{-i}]}x \succcurlyeq {g_{i}}_{[E_{-i}]}x.    
    \end{equation*}
\end{axiom}

\begin{table}[!htb]
    \begin{minipage}{.39\linewidth}
                \centering
          \begin{tabular}{c||c|c}
           $f_B$ &  $F$ &   $\neg F$   \\\hline\hline
        $B$   &  $y$ & $y$  \\\hline
        $\neg B$   &  $z$ & $z$  
    \end{tabular}
    \end{minipage}%
    \begin{minipage}{.26\linewidth}
                \centering
          $\succcurlyeq$
    \end{minipage}%
   \begin{minipage}{.36\linewidth}
                \centering
          \begin{tabular}{c||c|c}
           $f^\prime_B$ &  $F$ &   $\neg F$   \\\hline\hline
        $B$  &  $y^\prime$ & $y^\prime$  \\\hline
        $\neg B$   &  $z^\prime$ & $z^\prime$  
    \end{tabular}
    \end{minipage}%


\centering \quad $\iff$

    \begin{minipage}{.39\linewidth}
                \centering
          \begin{tabular}{c||c|c}
          ${f_B}_{[\neg F]}x$  &  $F$ &   $\neg F$   \\\hline\hline
        $B$  & $x$   &  $y$\\\hline
        $\neg B$   & $x$ &  $z$ 
    \end{tabular}
    \end{minipage}%
    \begin{minipage}{.23\linewidth}
                \centering
          $\succcurlyeq$
    \end{minipage}%
    \begin{minipage}{.36\linewidth}
                \centering
          \begin{tabular}{c||c|c}
            ${f^\prime_B}_{[\neg F]}x$ &  $F$ &   $\neg F$ \\\hline\hline
        $B$  & $x$ &  $y^\prime$ \\\hline
        $\neg B$   & $x$  &  $z^\prime$
    \end{tabular}
    \end{minipage}%
        \caption{Assuming that $[\neg F]$ is non-null, Subspace Separation requires that $f_B\succcurlyeq f^\prime_B$ if and only if ${f_B}_{[\neg F]}x \succcurlyeq {f^\prime_B}_{[\neg F]}x$. In the context of the insurance example, we can think of outcome $y,y'$ as the monetary coverage in case of a fire, whereas $z,z'$ as well as $x$ correspond to the insurance denying the claim.}\label{Table: illustration subspace separation}
\end{table}

\begin{theorem}\label{theorem: Full Independence}
    Assume that $\succcurlyeq$ exhibits dependence uncertainty and is MEU with prior set $\mathcal{C} \subseteq \mathcal{P}(\Omega; p_1, \ldots, p_n)$. The following are equivalent:
\begin{enumerate}
    \item $\succcurlyeq$ satisfies Subspace Separation.\label{item: Theorem full independence subspace separation}
    \item $\succcurlyeq$ is SEU with respect to the independent product $\pind = p_1 \otimes \ldots \otimes p_n$.\label{item: Theorem full independence SEU pind}
\end{enumerate}
\end{theorem}

Subspace Separation is strong enough to both fully eliminate dependence uncertainty, resulting in an SEU representation, and elicit the independent product as the considered prior. Theorem \ref{theorem: Full Independence} thus provides a decision-theoretic foundation for independence-based preferences under ambiguity, making behavioral phenomena such as correlation neglect testable.

\section{Dependence premia and comparative dependence aversion}
\label{section: comparative dependence uncertainty and dependence premia}

This section introduces two complementary measures of attitudes toward dependence. The first is the monetary compensation associated with replacing the decision maker's assessment of an act by its evaluation under the independent product: the \emph{dependence premium}. The second compares two decision makers by suggesting that if the one is less willing to choose a dependence-sensitive act over a dependence-neutral benchmark compared to the other, then this decision maker is \emph{more dependence averse}. Within the dependence-uncertainty model, the two measures induce the same ordering once risk preferences and marginal beliefs are fixed. Further, global non-negativity of dependence premia is equivalent to \emph{absolute dependence aversion}. We clarify the consequences for the MEU, smooth and variational model. 

\subsection{Dependence premia}

Consider a worker who receives labor income from firm $A$ and also owns shares in $A$. Suppose she can replace these shares by a security whose marginal return distribution is identical to that of the shares in $A$, but whose return is independent of her employment status. The replacement leaves the marginal distributions of employment and asset returns unchanged while eliminating their dependence. The amount of money that the worker is willing to pay for this replacement is a \emph{dependence premium}.

\begin{definition}[Dependence premium]
\label{definition: dependence premium}
Let $X=\mathbb R$, let $u\colon\mathbb R\to\mathbb R$ be continuous and strictly increasing, and suppose that $\succcurlyeq$ exhibits dependence uncertainty with respect to $u$ and marginals $p_1,\ldots,p_n$. The \emph{dependence premium} of $f\in\mathcal F$ is the unique number $\DP(f)$ satisfying
\begin{equation}\label{eq: dependence premium definition}
    \int_\Omega u\bigl(f-\DP(f)\bigr)\md\pind=V(f),
\end{equation}
where $(f-c)(\omega)=f(\omega)-c$.
\end{definition}

The premium is well defined and invariant under positive affine transformations of $u$. The switch to the independent product eliminates all non-independence and, as a side-effect, all dependence uncertainty. If $\DP(f)>0$, the decision maker is willing to give up $\DP(f)$ units of money to have $f$ evaluated under independence; if $\DP(f)<0$, she must be compensated for that change. Note that there can be a nonzero dependence premium even in the absence of ambiguity: an SEU decision maker with prior $p \neq \pind$ has a nonzero dependence premium function even though she faces no uncertainty about the joint distribution.

\subsection{Comparative dependence aversion}

Consider two workers who receive wage $K>0$ if they remain employed and zero otherwise. Each worker can purchase, at an up-front price $c>0$, a claim that pays a bonus $B>0$ if the firm succeeds and the worker remains employed. The two alternatives are shown in Table \ref{Table: worker bonus}.

\begin{table}[!htb]
\centering
\begin{minipage}{.48\linewidth}
\centering
\begin{tabular}{c||c|c}
$g$ & Success & Failure \\ \hline\hline
Employed   & $K$ & $K$ \\ \hline
Unemployed & $0$ & $0$
\end{tabular}
\end{minipage}
\hfill
\begin{minipage}{.48\linewidth}
\centering
\begin{tabular}{c||c|c}
$f$ & Success & Failure \\ \hline\hline
Employed   & $K-c+B$ & $K-c$ \\ \hline
Unemployed & $-c$ & $-c$
\end{tabular}
\end{minipage}
\caption{The status quo and the bonus claim.}
\label{Table: worker bonus}
\end{table}

The status quo is dependence neutral, whereas the attractiveness of the bonus claim relies on how firm success and continued employment depend on each other, as well as the attitude towards this dependence. Suppose the workers have the same marginal beliefs and risk preferences but may differ in their attitudes toward this dependence. If one worker accepts the claim while the other rejects it, it is natural to regard the latter as more dependence averse. More generally, whenever a more dependence-averse decision maker prefers a dependence-sensitive act to a dependence-neutral benchmark, a less dependence-averse decision maker should do so as well. This motivates the following definition, in the spirit of
\citet{yaari1969some}.

\begin{definition}[Comparative dependence aversion]
\label{definition: comparative dependence uncertainty aversion}
Let $\succcurlyeq$ and $\succcurlyeq^\prime$ be preference relations. We
say that $\succcurlyeq$ is \emph{more dependence averse} than
$\succcurlyeq^\prime$ if, for every $f\in\mathcal F$ and every
$g\in\FvNMdn$,
\[
f\succcurlyeq g \quad\Longrightarrow\quad f\succcurlyeq^\prime g.
\]
\end{definition}

\begin{remark}[Relation to comparative ambiguity aversion]
Definition \ref{definition: comparative dependence uncertainty aversion} is a dependence-specific formulation of the comparative ambiguity-aversion order of \citet{ghirardato2002ambiguity}. Their definition uses constant acts as benchmarks, whereas we use dependence-neutral acts. On the class of dependence-uncertainty preferences, however, the two definitions are behaviorally equivalent. The distinction within this domain is thus interpretive rather than behavioral. Dependence-neutral acts, unlike constant acts, retain marginal uncertainty while eliminating exposure to dependence. They are thus the natural benchmark for a comparative notion. Outside of the dependence-uncertainty model, the notions differ. Constant acts identify common risk preferences, while dependence-neutral acts additionally identify common marginal beliefs. The latter benchmark therefore has greater identifying content.
\end{remark}

\subsection{Representation consequences}

Proposition \ref{Proposition: Comparative dependence aversion} links the notion of a preference being more dependence averse than another to parameter properties of the dependence uncertainty representation, as well as the concept of dependence premia. Corollaries \ref{corollary: comparative DP meu and smooth} and \ref{Corollary: Absolute dependence aversion} derive its consequences for the MEU, smooth and variational models when considering the comparative as well as the logical absolute notion of dependence aversion.

\begin{proposition}\label{Proposition: Comparative dependence aversion}
Let $\succcurlyeq$ and $\succcurlyeq^\prime$ exhibit dependence uncertainty with representation parameters $(u,\alpha,p_1,\ldots,p_n)$ and $(u^\prime,\alpha^\prime,p_1^\prime,\ldots,p_n^\prime)$, respectively. The following are equivalent:
\begin{enumerate}[(i)]
    \item $\succcurlyeq$ is more dependence averse than $\succcurlyeq^\prime$.\label{item: comparative dependence def}

    \item The utility functions $u$ and $u^\prime$ are positive affine transformations of one another, $p_i=p_i^\prime$ for every $i$, and $\alpha(f)\geq\alpha^\prime(f)$ for every $f\in \mathcal{F}$.\label{item: comparative dependence functional}
\end{enumerate}
If, in addition, $X=\mathbb R$ and the utility functions are continuous and strictly increasing, the above conditions are also equivalent to
\begin{enumerate}[(i)]
    \setcounter{enumi}{2}
    \item $u$ and $u^\prime$ are positive affine transformations of one
    another, $p_i=p_i^\prime$ for every $i$, and
    $\DP(f)\geq\DP^\prime(f)$.
\end{enumerate}
\end{proposition}

\begin{corollary}[Comparative dependence aversion]\label{corollary: comparative DP meu and smooth}
Suppose $\succcurlyeq, \succcurlyeq^\prime$ exhibit dependence uncertainty. Then, the following statements hold.
\begin{enumerate}[(i)]
    \item\label{cor comp DP MEU} If in addition $\succcurlyeq, \succcurlyeq^\prime$ are MEU with prior sets $\mathcal C,\mathcal C^\prime$, then $\succcurlyeq$ is more dependence averse than $\succcurlyeq^\prime$ if and only if their utility functions are positive affinely related and $\mathcal C\supseteq\mathcal C^\prime$.\

    \item\label{cor comp DP smooth} If in addition $\succcurlyeq, \succcurlyeq^\prime$ are smooth with $\phi, \phi^\prime$ and the same second-order belief $\mu$. If $\mu$ is nondegenerate and $\phi,\phi^\prime$ are twice continuously differentiable, then $\succcurlyeq$ is more dependence averse than $\succcurlyeq^\prime$ if and only if their utility functions are positive affinely related and $\phi=h\circ\phi^\prime$ for some strictly increasing concave $h$. If $\mu $ is degenerate, both preferences are SEU and identical.

    \item\label{cor comp DP VP} If in addition $\succcurlyeq, \succcurlyeq^\prime$ are variational with minimal grounded cost functions $c^*,{c^*}^\prime$, then $\succcurlyeq$ is more dependence averse than $\succcurlyeq^\prime$ if and only if their utility functions are positive affinely related and, given a common utility normalization, ${c^*}(p)\leq {c^*}^\prime(p)$ for every $p\in\Delta(\Omega)$.
\end{enumerate}
\end{corollary}

Taking expected utility under the independent product as the natural dependence-neutral benchmark yields the absolute counterpart to the comparative notion in Definition \ref{definition: comparative dependence uncertainty aversion}.

\begin{definition}[Absolute dependence aversion]
\label{definition: absolute dependence aversion}
Let $\succcurlyeq_{\mathrm{ind}}$ denote the expected-utility preference generated by the decision maker's utility index $u$ and the independent product $\pind=\bigotimes_{i=1}^n p_i$ of her marginal beliefs. A preference relation $\succcurlyeq$ is \emph{absolutely dependence averse} if it is more dependence averse than $\succcurlyeq_{\mathrm{ind}}$ in the sense of Definition \ref{definition: comparative dependence uncertainty aversion}.
\end{definition}

The following corollary characterizes this property for MEU, smooth, and variational preferences and derives its implication for the dependence premium.

\begin{corollary}[Absolute dependence aversion]
\label{Corollary: Absolute dependence aversion}
Suppose that $\succcurlyeq$ exhibits dependence uncertainty with utility $u$ and marginals $p_1,\ldots,p_n$. Let $\succcurlyeq_{\ind}$ denote the SEU preference with utility $u$ and prior $\pind$. The following statements hold.
\begin{enumerate}[(i)]
    \item If $\succcurlyeq$ is MEU with prior set $\mathcal C$, then it is more dependence averse than $\succcurlyeq_{\ind}$ if and only if $\pind\in\mathcal C$.\label{item: Cor ADA MEU}
    \item If $\succcurlyeq$ is smooth, $\mu$ nondegenerate and $\phi$ twice continuously differentiable, then $\succcurlyeq$ is more dependence averse than $\succcurlyeq_{\ind}$ if and only if $\bar p_\mu:=\int p\md\mu(p)=\pind$ and $\phi$ is concave. If $\mu=\delta_p$, for arbitrary $\phi$, $\succcurlyeq$ is more dependence averse than $\succcurlyeq_{\ind}$ if and only if $p=\pind$.\label{item: Cor ADA smooth}
    \item If $\succcurlyeq$ is variational with minimal grounded penalty $c^*$, then it is more dependence averse than $\succcurlyeq_{\ind}$ if and only if $c^*(\pind)=0$.\label{item: Cor ADA VP}
\end{enumerate}

If, in addition, $X=\mathbb R$ and $u\colon\mathbb R\to\mathbb R$ is continuous and strictly increasing, then $\succcurlyeq$ is more dependence averse than $\succcurlyeq_{\ind}$ if and only if
\begin{equation*}
    \DP(f) \geq 0  \qquad\text{for every }f\in\mathcal F.
\end{equation*}
\end{corollary}

\section{Dependence invariance}\label{Section: dependence-invariance}

Dependence neutrality identifies acts whose expected utility is unaffected by dependence, see Section \ref{sec:dependence-neutral acts}. This section extends this idea by distinguishing acts with the same cardinal exposure to dependence as well as acts whose exposures differ only in scale. We provide characterizations of both. We then strengthen the independence axiom used in Theorem \ref{theorem: dependence uncertainty} and show that the two strengthenings restrict the otherwise act-dependent coefficient $\alpha$: it must be constant on the introduced equivalence classes. The result clarifies the connection between our model and Weak Certainty Independence and Certainty Independence.

\begin{definition}[Dependence equivalence]\label{definition: dependence equivalence}
For $f,g\in\mathcal F$, say that
\begin{enumerate}[(i)]
    \item $f$ and $g$ are \emph{cardinally dependence equivalent}, written
    $f\asymp g$, if
    \begin{equation*}\label{equation: cardinal dependence equivalence}
        \int_\Omega u(f)\,\md q=\int_\Omega u(g)\,\md q
        \qquad\text{for every affine }u\colon X\to\R
        \text{ and every }q\in Q,
    \end{equation*}
    \item $f$ and $g$ are \emph{ordinally dependence equivalent}, written
    $f\bowtie g$, if, for every affine $u\colon X\to\R$ and every
    $q,q'\in Q$,
    \begin{equation*}\label{equation: ordinal dependence equivalence}
    \begin{split}
        \int_\Omega u(f) \md q\geq \int_\Omega u(f)\,\md q'
        \quad\iff\quad
        \int_\Omega u(g)\md q\geq \int_\Omega u(g)\md q'.
    \end{split}
    \end{equation*}
    Because $Q$ is a vector space, $f \bowtie g$ is equivalent to $\int_{\Omega} u(f) \md q$ and $\int_{\Omega} u(g) \md q$ having the same sign for every affine $u\colon X\to\R$ and $q\in Q$.

\end{enumerate}
\end{definition}

Both relations are equivalence relations and $\asymp$ refines $\bowtie$. The class $[f]_{\asymp}$ contains all acts with exactly the same dependence exposure as $f$. Indeed, $f\asymp g$ means that, for every affine $u$, the difference $u(f)-u(g)$ of their utility profiles is dependence neutral. Thus, changing the dependence structure affects the expected utilities of $f$ and $g$ by the same amount. The coarser class $[f]_{\bowtie}$ contains all acts with the same directional exposure as $f$. If $f\bowtie g$, every change in dependence that raises expected utility under $f$ also raises it under $g$, and likewise for changes that lower expected utility. The magnitude of the change, however, may differ. Hence cardinal dependence equivalence preserves both the direction and magnitude of dependence exposure, whereas ordinal dependence equivalence preserves only its direction. For every constant act $x\in X$,
\begin{equation}
\label{equation: neutral equivalence class}
    [x]_{\asymp}=[x]_{\bowtie}=\FvNMdn.
\end{equation}

Table \ref{Table:example dependence invariance} illustrates the two relations.

\begin{table}[htb]
    \centering
    \begin{minipage}{.3\linewidth}
        \centering
        \begin{tabular}{c||c|c}
            $f$ & $U_2$ & $D_2$ \\\hline\hline
            $U_1$ & $x$ & $y$ \\\hline
            $D_1$ & $y$ & $y$
        \end{tabular}
    \end{minipage}
    \begin{minipage}{.3\linewidth}
        \centering
        \begin{tabular}{c||c|c}
            $g$ & $U_2$ & $D_2$ \\\hline\hline
            $U_1$ & $y$ & $y$ \\\hline
            $D_1$ & $y$ & $x$
        \end{tabular}
    \end{minipage}
    \begin{minipage}{.3\linewidth}
        \centering
        \begin{tabular}{c||c|c}
            $h$ & $U_2$ & $D_2$ \\\hline\hline
            $U_1$ & $\frac12x+\frac12 y$ & $y$ \\\hline
            $D_1$ & $y$ & $y$
        \end{tabular}
    \end{minipage}
    \caption{We have $\int_\Omega u(f) \md q
    =\int_\Omega u(g) \md q
    =2\int_\Omega u(h) \md q$ for all affine $u\colon X\rightarrow \mathbb{R}$ and $q\in Q$ since the sum of utilities on both diagonals is the same. Thus, $f$ and $g$ are cardinally dependence-equivalent since they have the same integral against every dependence structure, for every affine utility function. Further, all three acts are ordinally dependence-equivalent, since they order dependence structures in the same way, for every affine utility index.}
    \label{Table:example dependence invariance}
\end{table}

The following result characterizes the two relations. The common mixing coefficient in part \eqref{item: asymp mixture characterization} preserves the magnitude of the dependence exposure. Allowing different coefficients in part \eqref{item: bowtie mixture characterization} preserves only its positive ray.

\begin{proposition}\label{proposition: dependence equivalence characterization}
For all $f,g\in\mathcal F$:
\begin{enumerate}[(i)]
    \item\label{item: asymp mixture characterization}$f\asymp g$ if and only if there exist
    $h,h'\in\FvNMdn$ and $\beta\in(0,1]$ such that
    \begin{equation*}
        \beta f+(1-\beta)h=\beta g+(1-\beta)h'.
    \end{equation*}
    \item\label{item: bowtie mixture characterization} $f\bowtie g$ if and only if there exist
    $h,h'\in\FvNMdn$ and $\beta,\beta'\in(0,1]$ such that
    \begin{equation*}
        \beta f+(1-\beta)h=\beta' g+(1-\beta')h'.
    \end{equation*}
\end{enumerate}
\end{proposition}

Recall that the dependence-uncertainty representation evaluates an act according to $V(f)=\alpha(f)\min_{p\in\mathcal P}\int_\Omega u(f) \md p+(1-\alpha(f))\max_{p\in\mathcal P}\int_\Omega u(f) \md p$. In particular, $\alpha$ may vary freely across dependence-sensitive acts. Cardinally dependence-equivalent acts, however, differ only by a dependence-neutral profile component. Assigning them different coefficients would therefore make the decision maker's attitude toward dependence hinge on a component that is itself unaffected by dependence. The analogous argument applies to ordinally dependence-equivalent acts once invariance to a positive rescaling of dependence exposure is also imposed. It is consequently natural to require $\alpha$ to be constant on $\asymp$-classes, or, more strongly, on the coarser $\bowtie$-classes. The two independence axioms below deliver precisely these restrictions. They are the counterparts of Weak Certainty Independence \citep{maccheroni2006ambiguity} and Certainty Independence \citep{gilboa1989maxmin}, respectively, with constant acts replaced by dependence-neutral acts.

\begin{axiom}[Weak Dependence-neutral Independence]
\label{axiom: DN translation independence}
For all $f,g\in\mathcal F$, $h,h'\in\FvNMdn$, and $\lambda\in(0,1)$,
\begin{equation*}
\lambda f+(1-\lambda)h\succcurlyeq
      \lambda g+(1-\lambda)h \quad\Longleftrightarrow\quad
      \lambda f+(1-\lambda)h'\succcurlyeq
      \lambda g+(1-\lambda)h'.
\end{equation*}
\end{axiom}

\begin{axiom}[Dependence-neutral Independence]
\label{axiom: DN affine independence}
For all $f,g\in\mathcal F$, $h\in\FvNMdn$, and $\lambda\in(0,1)$,
\begin{equation*}
    f\succcurlyeq g
    \quad\Longleftrightarrow\quad
    \lambda f+(1-\lambda)h\succcurlyeq
    \lambda g+(1-\lambda)h.
\end{equation*}
\end{axiom}

Dependence-neutral Independence implies Weak dependence-neutral Independence.\footnote{This can be seen by observing that Axiom \ref{axiom: DN affine independence} is equivalent to $\lambda f + (1-\lambda) h \succcurlyeq \lambda g + (1-\lambda) h \iff \gamma f + (1-\gamma) h^\prime \succcurlyeq \gamma g + (1-\gamma) h^\prime $ for all $f,g\in \mathcal{F}$, $h, h^\prime \in \FvNMdn$, $\lambda, \gamma \in (0,1)$. Axiom \ref{axiom: DN translation independence} is the special case $\lambda = \gamma$.} Both axioms permit two dependence-sensitive acts with statewise indifferent consequences to be evaluated differently. The following consequentialism condition rules this out.

\begin{axiom}[Statewise consequentialism]\label{axiom: statewise consequentialism}
If $f(\omega)\sim g(\omega)$ for every $\omega\in\Omega$, then $f\sim g$.
\end{axiom}

\begin{theorem}[Cardinal and ordinal dependence invariance]
\label{theorem: dependence-invariant DU}
Suppose that $\succcurlyeq$ satisfies Axioms
\ref{axiom: non-trivial weak order}--\ref{axiom: dn dominance} and
Axiom \ref{axiom: statewise consequentialism}.  Then:
\begin{enumerate}[(i)]
    \item Axiom \ref{axiom: DN translation independence} holds if and only if $\succcurlyeq$ admits a dependence-uncertainty representation for which there is a function $\alpha_{\asymp}\colon\mathcal F/\!\asymp\,\to[0,1]$ such that
    \begin{equation}\label{equation: asymp invariant DU}
    \begin{split}
        V(f)={}&\alpha_{\asymp}([f]_{\asymp})
        \min_{p\in\mathcal P}\int_\Omega u(f)\md p\\
        &+\bigl(1-\alpha_{\asymp}([f]_{\asymp})\bigr)
        \max_{p\in\mathcal P}\int_\Omega u(f)\md p.
    \end{split}
    \end{equation}
    \item Axiom \ref{axiom: DN affine independence} holds if and only if $\succcurlyeq$ admits a dependence-uncertainty representation for which there is a function $\alpha_{\bowtie}\colon\mathcal F/\!\bowtie\,\to[0,1]$ such that
    \begin{equation}\label{equation: bowtie invariant DU}
    \begin{split}
        V(f)={}&\alpha_{\bowtie}([f]_{\bowtie})
        \min_{p\in\mathcal P}\int_\Omega u(f)\md p\\
        &+\bigl(1-\alpha_{\bowtie}([f]_{\bowtie})\bigr)
        \max_{p\in\mathcal P}\int_\Omega u(f)\md p.
    \end{split}
    \end{equation}
\end{enumerate}
In both cases, we have the same uniqueness properties as in Theorem \ref{theorem: dependence uncertainty}.
\end{theorem}
\section{Geometric properties of dependence}
\label{Section: Math properties of P}

The dependence uncertainty set $\mathcal{P}$ captures a DM's uncertainty about dependence between subspaces $\Omega_i$ in presence of given marginal distributions $p_i$. The dependence itself is distilled in the set $Q$ of dependence structures. This section provides geometric characterizations of $\mathcal{P}$ and relates dependence structures to the statistical concept of copulae. We first calculate the dimension of the dependence uncertainty set by deriving a spanning system of $Q$, provide characterizations of the extreme points of the convex polytope $\mathcal{P}$, and finally show that dependence structures generalize the concept of copulae to arbitrary product state spaces.

\subsection{Rectangular operations and the dimension of \texorpdfstring{$\mathcal{P}$}{P}}

Recall that the set of dependence structures, $Q$, is defined as the set of all signed measures $q \colon 2^\Omega \to \R$ such that $q([\omega_i]) = 0$ for all $i \in \{1, \ldots, n\}$ and $\omega_i \in \Omega_i$. Embedding $Q$ in $\R^{\#\Omega}$ (or, equivalently, scaling and adding the values point-wise), it is straightforward to see that $Q$ bears the structure of a finite-dimensional vector space over the reals. Restricting the state space to the Cartesian product of the supports of the $p_i$, the independent product $\pind$ is an interior point of $\mathcal{P}$ and we can add any (sufficiently scaled) $q \in Q^* := Q(\supp(p_1) \times \ldots \times \supp(p_n))$ to it, cf.\ Proposition \ref{Proposition: decomposition}. In that case, the size of $\mathcal{P}$ is measured by the vector space dimension of $Q^*$, i.e., $\dim(\mathcal{P}) := \dim_{\R}(Q^*)$. Notably, $\dim \mathcal{P}$ does not depend on the precise values of the marginals $p_i$, as long as the cardinality of their supports is the same.

In Example \ref{Example: dual sourcing}, the marginals have full support and $q = (\begin{smallmatrix}
    a &-a\\
    -a & a
\end{smallmatrix})$ is, up to a scalar, the only dependence structure of the two-by-two state space. Consequently, $\dim(\mathcal{P}) = 1$, as indicated by the single parameter $a$.

To understand the set of dependence structures for a general state space $\Omega = \Omega_1 \times \ldots \times \Omega_n$, we define the following family of dependence structures that mimic the observations from the introductory example.
\begin{definition}\label{Definition: rectangular Operation}
    Let $i \in \{1, \ldots, n\}$ and $\omega_i^* \neq \omega_i^\dagger$ in $\Omega_i$, $\omega_{-i}^* \neq \omega_{-i}^\dagger$ in $\Omega_{-i}$. A dependence structure $q$ with $q(\omega_i^*, \omega_{-i}^*) = q(\omega_i^\dagger, \omega_{-i}^\dagger) = a$, $q(\omega_i^*, \omega_{-i}^\dagger) = q(\omega_i^*, \omega_{-i}^\dagger) = -a$, for any $0 \neq a \in \R$, and $q(\omega) = 0$ otherwise, is called a \emph{rectangular operation}.
\end{definition}
A rectangular operation is supported only on four states that can be thought of as rectangle and shifts probability mass $2a$ from one diagonal to the other, see Figure \ref{Figure: rectangular operation definition}.

\begin{figure}[htb]
    \centering
    \begin{tikzpicture}
        \coordinate (A) at (0,0);
        \coordinate (B) at (3,0);
        \coordinate (C) at (0,1.5);
        \coordinate (D) at (3,1.5);

        \draw (A) node[]{$(\omega_i^\dagger, \omega_{-i}^*)$};
        \draw (B) node[]{$(\omega_i^\dagger, \omega_{-i}^\dagger)$};
        \draw (C) node[]{$(\omega_i^*, \omega_{-i}^*)$};
        \draw (D) node[]{$(\omega_{i}^*, \omega_{-i}^\dagger)$};

        \draw[<-] (C)++(1,0)--($(D) + (-1,0)$)node[midway,above]{$a$};
        \draw[->] (A)++(1,0)--($(B) + (-1,0)$)node[midway,above]{$a$};

        \draw[dotted] (C)++(0.8,-0.4) -- ($(B)+(-0.8,0.4)$);
        \draw[dotted] (A)++(0.8,0.4) -- ($(D)+(-0.8,-0.4)$);
        
    \end{tikzpicture}
    \caption{Visualization of a rectangular operation from Definition \ref{Definition: rectangular Operation}. A total mass of $2a$ is shifted from one diagonal to the other.}
    \label{Figure: rectangular operation definition}
\end{figure}

Indeed, any dependence structure can be written as a linear combination of rectangular operations, aiding to find a basis and thus yielding the dimension of $\mathcal{P}$.
\begin{proposition}\label{Proposition: Dimension of P}
Rectangular operations form a spanning system of $Q$. Setting $m_i := \# \supp(p_i)$, the dimension of $\mathcal{P}$ is equal to
 $\prod\limits_{i=1}^n m_i - \sum\limits_{i=1}^n (m_i-1)  -1$.
\end{proposition}
The intuition behind the dimension of $\mathcal{P}$ is straightforward: There are $\prod_{i=1}^n \# m_i$ many variables, represented by the values of a probability $p \in \mathcal
P$, minus $1$ for being a probability. For each $i \in \{1, \ldots, n\}$, there are $m_i$ many marginal restrictions, one of them being redundant as the marginals are probabilities.

\subsection{Extreme points of dependence}\label{Section: Extreme Points}
The set $\mathcal{P}$ is a compact convex polytope and thus the convex hull of its finite set of extreme points, which we denote by $\ext \mathcal{P}$.\footnote{By definition, a compact convex polytope is the convex hull of finitely many points in a Euclidean space. Equivalently, it is the intersection of finitely many half-spaces \citep{grunbaum2003convex}. Since $\mathcal{P}$ is the intersection of the compact convex polytope $\Delta(\Omega)$ and the half-spaces encoding the marginal constraints, it is a compact convex polytope as well.} In the following, we provide two equivalent characterizations the extreme points of $\mathcal{P}$. Firstly, as probabilities with minimal support. Secondly, as the local maximizers of certain classes of divergences, used in information theory. Furthermore, we elaborate how dependence structures map into the statistical object of a copula, establishing them as the generalization to un-ordered spaces.

\begin{definition}\label{Definition: p maximally zero}
    A probability vector $p \in \mathcal{P}$ is called \emph{maximally zero}, if for any $p' \in \mathcal{P}$ with $p(\omega) = 0 \Rightarrow p'(\omega) = 0$ for all $\omega$ we have $p = p'$.
\end{definition}
In other words, $p \in \mathcal{P}$ is maximally zero if and only if the only $p' \in \mathcal{P}$ that is absolutely continuous w.r.t.\ $p$, i.e., $p' \ll p$, is $p'=p$. %
A maximally zero probability vector $p$ assigns probability $0$ to as many states $\omega \in \Omega$ as possible in the sense that there is no other probability $p' \in \mathcal{P}$ that vanishes in the same states as $p$.\footnote{Defining the \emph{capacity} $\upsilon(E) := \min_{p \in \mathcal{P}} p(E)$, one can easily identify a maximally zero $p$ as an event $E := p^{-1}(0)$ with the property $\upsilon(E) = 0$, $\upsilon(E')>0$ for all $E \subsetneq E'$. Furthermore, the \emph{core} of $\upsilon$ is equal to $\mathcal{P}$, implying that the capacity is \emph{exact} \citep{schmeidler1972cores}. However, $\upsilon$ is not convex, thus, the \emph{Choquet expected utility} \citep{schmeidler1989subjective} w.r.t.\ $\upsilon$ does not agree with the respective maxmin expected utility \citep{gilboa1989maxmin}.} One can equivalently think of probabilities having minimal support. The intuition behind maximally zero probability vectors being extreme points is that an element in the interior resp.\ on a face of $\mathcal{P}$ can be further pushed to a face of lower index of the simplex, thereby creating another zero. This is formally shown in Theorem \ref{Theorem: extreme points via mutual information}.\bigskip

The extreme points of $\mathcal{P}$ can also be linked to the concept of \emph{divergences} in statistics and information geometry. Intuitively, a divergence measures the distance between two probability distributions. We are going to see that the local maximizers of certain classes of divergence towards the independent product are exactly the extreme points of $\mathcal{P}$. To this end, we briefly define the two concepts of $f$-divergence and Bregman divergence for finite state spaces, cf.\ \cite{amari2016information}: Firstly, for a convex function $f \colon \R_{\geq 0} \to \R$, the \emph{$f$-divergence} from $p$ to $p'$ is defined as $\Divergence{f}{p}{p'} := \sum\limits_{\omega \in \Omega} f \left( \frac{p(\omega)}{p'(\omega)}\right) \cdot p'(\omega)$. Second, for a continuously differentiable, strictly convex function $F \colon [0,1]^N \to \R$, the \emph{Bregman divergence} is defined as $\Divergence{F}{p}{p'} := F(p) - F(p') - \langle \nabla F(p'), p-p'\rangle$, where $p,p'$ are interpreted as vectors in $[0,1]^N$.
\begin{definition}
    We call $D \colon \Delta(\Omega) \times \Delta(\Omega) \to \R$ a \emph{divergence with infinitely decreasing slope in $0$}, if it is a strictly increasing transformation of either an $f$-divergence, where $f$ is differentiable with $\lim_{t \to \infty}f'(t) = -\infty$, or a Bregman divergence with $\lim_{\vec{t} \to 0}\frac{\partial}{\partial \omega}F(\vec{t}) = -\infty$ for all $\omega \in \Omega$.
\end{definition}
\begin{example}
    The following well-known $f$-divergences used in information theory have infinitely decreasing slope in $0$: The $\alpha$-divergence for $\alpha \in [0,1]$, including (generalized, reversed) Kullback-Leibler divergence, Jeffrey’s divergence, Jensen-Shannon divergence, squared Hellinger distance, Pearson $\chi^2$-divergence (aka reverse Pearson). Furthermore, the Itakura–Saito divergence as a Bregman divergence and the Rényi divergence for $\alpha \in [0,1]$, as a strictly increasing transformation of an $\alpha$-divergence, have infinitely decreasing slope in $0$.
\end{example}

The following theorem ties the concept of divergencies from information theory and maximally zero probability vectors to the geometric property of extreme points of dependence uncertainty.

\begin{theorem}\label{Theorem: extreme points via mutual information}
    The following are equivalent for a probability $p \in \mathcal{P}$ and any divergence $\Divergence{}{\cdot}{\cdot}$ with infinitely decreasing slope in $0$:
    \begin{enumerate}[(i)]
        \item $p$ is an extreme point of $\mathcal{P}$.\label{item: p extreme}
        \item $p$ is maximally zero.\label{item: p maximally zero}
        \item $p$ is a strict local maximizer of the divergence $\Divergence{}{\cdot}{\pind}$.\label{item: p local maximizer of mutual information}
    \end{enumerate}
\end{theorem}
\subsection{Relation to copulae}\label{Sub Sec: copulae}%

The concept of dependence structures $q$ defined in this article is applicable for probability distributions on arbitrary state spaces. If multivariate real-valued distributions are available, dependence is typically modeled by \emph{copulae}, in particular in statistics. In the following, we explain that and how these two concepts map to one another on a real state space. To this end, let $\Omega = \Omega_1 \times \ldots \times \Omega_n$ where $\{1\} \subseteq \Omega_i \subseteq (0,1]$ are closed.\footnote{Thus, we allow $\Omega$ to be finite, but do not require it. For a probability $p$ on $\Omega$, we generally define dependence structures as signed measures for which all marginals are constantly $0$.} Since copulae require a zero in their domain to be mapped to zero, and we would not be able to handle point masses on zero, we need to include them to get coinciding concepts by setting $\hat{\Omega}_i := \Omega_i \cup \{0\}$ and $\hat{\Omega} = \hat{\Omega}_1 \times \ldots \times \hat{\Omega}_n$. A \emph{subcopula} is a function $C \colon \hat{\Omega} \to [0,1]$ such that $C(\omega_1, \ldots, \omega_n) = 0$ if at least one $\omega_i = 0$, $C(1, \ldots, 1, \omega_i, 1, \ldots, 1) = \omega_i$ and $C$ is $n$-increasing, i.e., the \emph{$C$-volume}
\begin{equation*}
    V_C( \{ \underline{\omega_1}, \overline{\omega_1} \} \times \ldots \times \{ \underline{\omega_n}, \overline{\omega_n} \}) := \sum_{\omega \in \{ \underline{\omega_1}, \overline{\omega_1} \} \times \ldots \times \{ \underline{\omega_n}, \overline{\omega_n} \}} C(\omega) \cdot (-1)^{\# \{ i \mid \omega_i = \underline{\omega_i} \}} 
\end{equation*}
is non-negative for all Cartesian products of $\underline{\omega_i} \leq \overline{\omega_i}$, cf.\ \cite{nelsen2006introduction}. A subcopula with full range $\hat{\Omega} = [0,1]^n$ is called a \emph{copula}. Let $\Cop$ denote the set of all subcopulae on $\hat{\Omega}$, which is a convex set. The \emph{independent} (sub-)copula is defined by $\Cind (\omega_1, \ldots, \omega_n) = \omega_1 \cdot \ldots \cdot \omega_n$.

Denote by $\pind := p_1 \otimes \ldots \otimes p_n$ the independent product of the ``order-uniform'' marginals on $\Omega_i$, defined by $p_i([0,\omega_i] \cap \Omega_i) = \omega_i$. The following proposition establishes the structural equivalence of copulae and dependence structures on real state spaces. The formal derivation is detailed in the appendix.

\begin{proposition}\label{Proposition: copula}
    The set of copulae is isomorphic as a convex space to the set $\{ q \in Q \mid \pind + q \in \Delta(\Omega)\}$ of dependence structures.
\end{proposition}

Consequently, copulae and dependence structures are two exchangeable lenses through which dependence can be described. Note that the above isomorphism carries over if copulae and dependence structures are used to decompose a joint distribution with non-uniform marginals via the usual probability integral transform.

We again highlight that dependence structures do not rely on distribution functions---and thus not on an ordered state space---as copulae do. Furthermore, dependence structures suggest a natural addition of copulae: Setting $C_1 \oplus C_2 := C^{q^{C_1} + q^{C_2}} = C_1 + C_2 - \Cind$ whenever $q^{C_1} + q^{C_2} \in \{ q \in Q \mid \pind + q \in \Delta(\Omega)\}$ allows to combine dependencies.

As a concrete illustrative example, consider the copula $C$ and the dependence structure $q$ depicted in Table \ref{Table: copula dependence structure example}. Note that $C$ (resp.\ $C - \Cind$) can be obtained from $p = \pind +q$ (resp.\ $q$) by summing up the values in resp.\ rectangles starting in the top left corner.
\begin{table}[htb]
    \centering
    \quad
    \begin{tabular}{c||cc}
        $C$ & $\frac{2}{3}$ & $1$ \\\hline\hline
        $\frac{1}{4}$  &  $\frac{1}{6} + x$ & $ \frac{1}{4}$\\
        $1$  & $\frac{2}{3}$ & $1$
    \end{tabular}
    \qquad\quad
    \begin{tabular}{c||cc}
        $C - C_{\ind}$  & $\frac{2}{3}$ & $1$ \\\hline\hline
        $\frac{1}{4}$ &  $x$ & $0$\\
        $1$ & $0$ & $0$
    \end{tabular}
    \\\smallskip
    \begin{tabular}{c||cc}
        $\pind + q$ & $\frac{2}{3}$ & $1$ \\\hline\hline
        $\frac{1}{4}$ & $\frac{1}{6}+x$   & $\frac{1}{12} -x$\\
        $1$ &   $\frac{1}{2} -x$ & $\frac{1}{4}+x$
    \end{tabular}
    \qquad
    \qquad
    \begin{tabular}{c||cc}
        $q$ & $\frac{2}{3}$ & $1$ \\\hline\hline
        $\frac{1}{4}$ & $x$  & $-x$\\
        $1$ &   $-x$ & $x$
    \end{tabular}
    \caption{A copula $C$ and its corresponding dependence structure $q$ for $\Omega_1 = \{ \frac{1}{4}, 1 \}$, $\Omega_2 = \{ \frac{2}{3}, 1 \}$ with $x \in [-\tfrac{1}{6}, \tfrac{1}{12}]$.}
    \label{Table: copula dependence structure example}
\end{table}

\paragraph{Acknowledgements} %
The authors thank Yves Breitmoser, Rabanus Derr, Jürgen Eichberger, Fabian Fuchs, Itzhak Gilboa, Ani Guerdjikova, Paul Heidhues, Edith Hübner, Bernhard Kasberger, Peter Klibanoff, Eva Lütkebohmert-Holtz, Fabio Maccheroni, Lasse Mononen, Max Nendel, Georg Nöldeke, Pietro Ortoleva, Jurek Preker, Carmen Rähmisch, Stefan Richter, Frank Riedel, Alessandro Sgarabottolo, Jean-Marc Tallon, the participants of the 2nd Workshop on Learning Under Weakly Structured Information at the Tübingen AI Center, the Workshop on Unforeseen Contingencies in Grenoble, the SAET conference 2025 in Ischia, the FUR conference 2026 in Alicante, the Swiss Theory Day 2026 in Zürich, the research seminars in Bern and Basel, and an anonymous referee of the ISIPTA 2025 conference for helpful comments and fruitful discussions. We are particularly grateful for mutual research stays at Basel and Bielefeld made possible by Georg and Frank.

\setlength{\bibsep}{0pt}
\bibliography{literature}

\appendix
\section{Proofs}

\subsubsection*{Proofs of Section \ref{Section: P and Q}}
\begin{proof}[Proof of Proposition \ref{Proposition: decomposition}]
    Note that $\mathcal{P}(\Omega; p_1, \ldots, p_n)$ identifies with the intersection of the simplex $\Delta(\Omega)$ and the solution set to the inhomogeneous linear equation system $\sum\limits_{\omega_{-i} \in \Omega_{-i}} p(\omega_i,\omega_{-i}) = p _i(\omega_i)$ for all $i \in \{1, \ldots, n\}$ and $\omega_i \in \Omega_i$. To conclude, observe that $\pind$  is a particular solution to the inhomogeneous equation system while $Q$ is the solution set to the homogeneous counterpart.
\end{proof}

\subsubsection*{Proofs of Section \ref{sec:dependence-neutral acts}}
\begin{proof}[Proof of Proposition \ref{Proposition: characterization dependence neutrality for all notions}]
    Sufficiency is straightforward and thus omitted.\\
    Let $f \in \FvNMdn$ be dependence neutral. For $\omega \in \Omega$ let $f(\omega)(z)$ denote the probability that lottery $f(\omega)$ assigns to a prize $z \in Z$. Let $Z_f := \{ z \in Z \mid \exists \omega \colon f(\omega)(z) > 0 \}$ be the finitely many prizes that are assigned a positive mass under $f$ under some $\omega$. Since $f \in \FvNMdn$ and choosing the linear utility function, induced by $z \mapsto \one_{z^*}(z)$ on $\Delta(Z)$ for $z^* \in Z$, we have $(*)\, \int_{\Omega} f(\omega)(z) \md q = 0$ for all $q \in Q$ and $z \in Z$. For each $z \in Z_f$, let $\omega^z \in \argmin_{\omega \in \Omega} f(\omega)(z)$ and $c^z := f(\omega^z)(z)$. For each $i =1, \ldots, n$ we define the auxiliary functions $g_i \colon \Omega \times Z \to [0,1]$ by $g_i(\omega)(z) := f(\omega_i, \omega_{-i}^z) - c^z$ for $z \in Z_f$ and $0$ for $z \notin Z_f$. %
    We claim $f(\omega)(z) = c^z + \sum_{i=1}^n g_i(\omega)(z)$ for all $\omega \in \Omega, z \in Z$. To see this, fix any $z \in Z_f$ and consider the difference function $\omega \mapsto f(\omega)(z) - c^z - \sum_{i=1}^n g_i(\omega)(z)$. We show that it is constant and equal to $0$. First, plugging in $\omega = \omega^z$, we find $f(\omega^z) - c^z - \sum_{i=1}^n g_i(\omega^z)(z)=c^z -  c^z - 0 = 0$. Second, we show that changing one coordinate at a time does not alter the difference, thus proving that the difference is constant. To this end, consider for any fixed $j \in \{1, \ldots, n\}$ two states $\omega = (\omega_j, \omega_{-j})$ and $\omega' = (\omega_j', \omega'_{-j})$ with $\omega_j \neq \omega_j'$, $\omega_{-j} = \omega_{-j}'$ and some $\omega_{-j} \in \Omega_{-j}$. We find
    \begin{align*}
         &\left(f(\omega)(z)-c^z - \sum_{i=1}^n g_i(\omega)(z)\right) - \left(f(\omega')(z)-c^z - \sum_{i=1}^n g_i(\omega')(z)\right)\\
         =& f(\omega)(z) - f(\omega')(z) + \sum_{i=1}^n \left(f(\omega_i', \omega_{-i}^z)(z) - f(\omega_i, \omega_{-i}^z)(z) \right)\\
          =& f(\omega_j,\omega_{-j})(z) - f(\omega'_j, \omega_{-j})(z) + f(\omega_j', \omega_{-j}^z)(z) - f(\omega_j,\omega_{-j}^z)(z) = 0,
    \end{align*}
    where the last equality follows directly if $\omega_{-j} = \omega_{-j}^z$, or otherwise by applying $(*)$ to the rectangular operation defined on $\omega, \omega', (\omega_i, \omega_{-i}^z), (\omega_i', \omega_{-i}^z)$. %
    As a next step, we claim that for each $i=1, \ldots, n$ the function $\omega \mapsto \sum_{z \in Z_f}g_i(\omega)(z)$ is constant. To see this, take again $\omega = (\omega_j, \omega_{-j}), \omega' =  (\omega'_j, \omega_{-j}') \in \Omega$ with $\omega_i \neq \omega_i'$ and $\omega_{-j} = \omega_{-j}'$. If $i \neq j$, we have $g_i(\omega)(z)-g_i(\omega')(z) = f(\omega^z)(z) - f(\omega^z)(z) = 0$ for any $z \in Z_f$ and thus $\sum_{z \in Z_f} g_i(\omega)(z) - \sum_{z \in Z_f} g_i(\omega')(z) = 0$. If $j=i$, we find $g_i(\omega)(z) - g_i(\omega')(z) = f(\omega_i, \omega_{-i}^z)(z) - f(\omega'_i, \omega_{-i}^z)(z) = f(\omega)(z) - f(\omega')(z)$, using again $(*)$ if $\omega_{-i} \neq \omega_{-i}^z$. Consequently,
    \begin{align*}
        \sum_{z \in Z_f} g_i(\omega)(z) - \sum_{z \in Z_f} g_i(\omega')(z) &= \sum_{z \in Z_f} (g_i(\omega)(z) - g_i(\omega')(z))\\
        &= \sum_{z \in Z_f} (f(\omega)(z) - f(\omega_i')(z))\\
        &= \sum_{z \in Z_f} f(\omega)(z) - \sum_{z \in Z_f} f(\omega_i')(z) = 1 - 1 = 0.
    \end{align*}
    Denote these constants by $\hat{\gamma}_i$ and note $\hat{\gamma}_i \geq 0$. Finally, for $i=1, \ldots, n$, define $f_i(\omega)(z) := \hat{\gamma}_i^{-1} \cdot g_i(\omega)(z)$ if $\gamma_i > 0$ and $f_i(\omega)= \one_{z^*}$ for some fixed prize $z^*$ if $\gamma_i = 0$. Let $\hat{\gamma}_{n+1}:=\sum_{z \in Z_f} c^z$. If $\hat{\gamma}_{n+1} = 0$, define the constant lottery $x(z) := \one_{z^*}$, otherwise, let $x(z') := \hat{\gamma}_{n+1}^{-1} \cdot \sum_{z \in Z_f} c^{z} \one_{z}(z') > 0$. Note that $f_i \in \mathcal{F}_i$, $x \in X$ and $f = \hat{\gamma}_{n+1}x + \sum_{i=1}^n \hat{\gamma}_i f_i$ by construction. As the linear combination of acts/lotteries is an act itself, $\sum_{i=1}^{n+1} \hat{\gamma}_i = 1$. If $\hat{\gamma}_{n+1}>0$, one can arbitrarily redistribute its weight together with corresponding parts $c^z$ to the $f_i$ acts, making the desired decomposition unique up to this re-distribution.
\end{proof}

\begin{proof}[Proof of Proposition \ref{Proposition: Fudn adjusted to marginals}]
    For the first assertions let $f \in \Fudn(\Omega)$ and $ q^\circ \in Q(\Omega^\circ)$ be arbitrary. By setting $q(\omega) =q^\circ(\omega)$ if $\omega \in \Omega^\circ$ and $0$ otherwise, we obtain $q \in Q(\Omega)$. Thus, $0 = \int_{\Omega}u(f) \md q = \int_{\Omega^\circ}u(f) \md q = \int_{\Omega^\circ} u(f) \md q^\circ = \int_{\Omega^\circ} u(f\res_{\Omega^\circ}) \md q^\circ$, i.e., $f \res_{\Omega^\circ} \in \Fudn(\Omega^\circ)$ and hence $f \in \FudnO(\Omega; p_1, \ldots, p_n)$.\\
    For the equivalence, let first $f \in \FudnO(\Omega; p_1, \ldots, p_n)$ and $p \in \mathcal{P}(\Omega; p_1, \ldots, p_n)$ be arbitrary. Note that $\pind^{-1}(0)\subseteq \pind^{-1}(0)$. Consequently, $\int_{\Omega}u(f) \md p = \int_{\Omega^\circ} u(f \res_{\Omega^\circ}) \md p \res_{\Omega^\circ} = \int_{\Omega^\circ} u(f \res_{\Omega^\circ}) \md \pind \res_{\Omega^\circ} = \int_{\Omega}u(f) \md \pind$.\\
    For the other direction, let $f \in \mathcal{F}$ be with $f \res_{\Omega^\circ} \notin \Fudn(\Omega^\circ)$. Then there is a $q^\circ \in Q^\circ$ with $\int_{\Omega^\circ} u(f\res_{\Omega^\circ}) \md q^\circ \neq 0$. Since $\pind \res_{\Omega^\circ}$ is a relative interior point of $\Omega^\circ$, there is an $\epsilon>0$ such that $p^\circ := \pind \res_{\Omega^\circ} + \epsilon q^\circ$ is a probability measure on $\Omega^\circ$ with the same marginals as $\pind \res_{\Omega^\circ}$. Extend $p^\circ$ to a probability measure $p \in \mathcal{P}(\Omega; p_1, \ldots, p_n)$ by setting it $0$ outside of $\Omega^\circ$. It follows that $\int_{\Omega} u(f) \md (\pind-p) = \int_{\Omega^\circ} u(f \res_{\Omega^\circ}) \md (\pind \res_{\Omega^\circ} - p^\circ) = \epsilon \int_{\Omega^\circ}u(f \res_{\Omega^\circ}) \md q^\circ \neq 0$, so that the expectation of $f$ under $u$ over $\mathcal{P}(\Omega; p_1, \ldots, p_n)$ is not constant.
\end{proof}

\subsubsection*{Proofs of Section \ref{sec:axioms and representation}}
In preparation of the proof of Theorem \ref{theorem: dependence uncertainty}, we state auxiliary results that construct to any $f \in \mathcal{F}$ lower and upper sleeves $\underline{f},\overline{f} \in \FvNMdn$ with $\underline{f} \leq f \leq \overline{f}$. %
These are constructed iteratively in an algorithm. The first two results make use of the finite set of extreme points $\ext \mathcal{P}$ of the compact convex polytope $\mathcal{P}$, cf.\ Section \ref{Section: Extreme Points}. The first lemma tells us where we can safely change $f$.

\begin{alemma}\label{ALemma: argmin-ex-P set and stars}
    Let $u \colon X \to \R$ and $f \in \mathcal{F}$ be arbitrary. Define $P^* := \{ p^* \in \ex \mathcal{P} \mid p^* \in \argmin_{p \in \mathcal{P}} \int_{\Omega} u \circ f \md p \}$ and let $S := \bigcap_{p^* \in P^*} {p^{*}}^{-1}(0)$, $\mathcal{O}:=\bigcap_{p \in \mathcal{P}} p^{-1}(0)$. Then $S = \mathcal{O}$ if and only if $P^* = \ex(\mathcal{P})$.
    \begin{proof}[Proof of Lemma \ref{ALemma: argmin-ex-P set and stars}]
        Let first $P^* = \ex \mathcal{P}$. Since $\pind \in \conv \ex(\mathcal{P}) = \conv P^*$ and has support $\mathcal{O}$, $S = \mathcal{O}$. Conversely, let $S = \mathcal{O}$. Then, the probability $\hat{p} := \sum_{p^* \in P^*} \tfrac{1}{\# P^*} \cdot p^* \in \conv P^*$ has support $\mathcal{O}$. Now assume by means of contradiction that there is a $\tilde{p} \in \ex(\mathcal{P}) \setminus P^*$. Thus, $\min_{p \in \mathcal{P}} \int_{\Omega} u \circ f \md p = \int_{\Omega} u \circ f \md \hat{p} < \int_{\Omega} u \circ f \md \hat{p}$. For $\epsilon >0$ sufficiently small, the probability $\hat{p} - \epsilon (\hat{p}-\tilde{p}) \in \mathcal{P}$ and has strictly lower expected utility than $\hat{p}$,  a contradiction.
    \end{proof}
\end{alemma}

The next proposition produces an act with the desired utility profile.

\begin{aproposition}[Algorithm step]\label{AProposition: algorithm to find rectangular equivalent}
    Let $u \colon X \to \R$ and $f \in \mathcal{F}$ be arbitrary. Then, there is $h \in \Fudn$ with $h \leq f$ and $\min_{p \in \mathcal{P}} \int_{\Omega} u \circ f \md p = \min_{p \in \mathcal{P}} \int_{\Omega} u \circ h \md p$.
    \begin{proof}[Proof of Proposition \ref{AProposition: algorithm to find rectangular equivalent}]
        Define $P^*:= \{ p ^* \in \ex(\mathcal{P}) \mid p^* \in \argmin_{p \in \mathcal{P}} \int_{\Omega} u \circ f \md p \}$. If $P^* = \ex(\mathcal{P})$, we are done, as then every $p$ is a minimizer and yields the same expected payoff, i.e., we can set $h := f$. Otherwise, iteratively change $f$ as follows: As $P^* \neq \ex(\mathcal{P})$, we have $S := \bigcap_{p^* \in P^*} {p^*}^{-1}(0) \neq \mathcal{O}:= \bigcap_{p \in \mathcal{P}} p^{-1}(0)$ by Lemma \ref{ALemma: argmin-ex-P set and stars}. Set $\hat{p} := \sum_{p^* \in P^*} \tfrac{1}{\# P^*} \cdot p^*$ which is an element of maximal support in $\conv P^*$ and fulfills $\hat{p}^{-1}(0) = S \neq \mathcal{O}$. Now, consider $\{ \tilde{p} - \hat{p} \mid \tilde{p} \in \ex(\mathcal{P}) \setminus P^* \} \subseteq Q$ and let $q$ be one of its minimal elements w.r.t.\ $\int_{\Omega} u \circ f \md q$, which indeed is positive as otherwise the corresponding $\tilde{p}$ was in $P^*$. %
        Therefore, $\int_{\Omega} \one_{q > 0} u \circ f \md q > - \int_{\Omega} \one_{q < 0} u \circ f \md q$. By its minimality, $\hat{p}$ thus has a zero in a state $\omega^* \in q^{-1}((0,\infty))$. By Unboundedness, i.e., Axiom \ref{axiom: unboundedness}, and linearity of $u$ on the mixture space $X$, there exists $x \in X$ such that $u(x) = -\int_{\Omega \setminus \{ \omega^*\}} u \circ f \md q$. Defining $h(\omega^*) = x$ and $h(\omega) = f(\omega)$ for $\omega \neq \omega^*$, we have $\int_{\Omega} u \circ h \md q = 0$. Note that the evaluation of $h$ under any $p^* \in P^*$ is the same as under $f$ and equal to the one under $\tilde{p} = \hat{p}+q$. That these are indeed minimizers for $h$ can be seen as follows: Assume by means of contradiction that there is $p' \in \ex(\mathcal{P})$ with strictly smaller evaluation. Then $\int_{\Omega}u \circ f \md (p'-\hat{p})>0$, but $\int_{\Omega} u \circ h \md (p'-\hat{p})<0$, violating $q$'s minimality.
    \end{proof}
\end{aproposition}

Finally, the following lemma turns the above act into one in $\FvNMdn$.

\begin{alemma}\label{lemma: Fudn to FvNMdn}
    Let $\succcurlyeq$ be a preference relation on $\mathcal{F}$. Let $u:X \rightarrow \mathbb{R}$ be affine and reflect the risk preference of $\succcurlyeq$. Let $f\in \Fudn$. Then there exists an $h\in \FvNMdn$ such that $f(\omega) \sim h(\omega)$ for all $\omega\in \Omega$. 
\begin{proof}[Proof of Lemma \ref{lemma: Fudn to FvNMdn}]
    Let $u:X\rightarrow \mathbb{R}$ be affine and $f\in \Fudn$. If $u(f)$ is constant, then simply choose $h \equiv x$ with $u(x)=u(f(\omega))$ for some $\omega\in \Omega$. So assume that $u(f)$ is not constant. Without loss we can assume that $\min_{\omega\in \Omega} u(f(\omega))=0$ and $\max_{\omega\in \Omega} u(f(\omega))=1$. This is possible since $u$ is unique up to positive affine transformations. Consider consequences $x\in \arg\max_{\omega\in \Omega} u(f(\omega))$ and $y\in \arg\min_{\omega\in \Omega} u(f(\omega))$, i.e., $u(x)=1$ and $u(y)=0$. 
    
    We construct the act $h$ by assigning to each state a consequence which is a lottery of the kind $\beta x + (1-\beta)y$. For $\omega\in \Omega$, define $h(\omega) = \beta(\omega) x + (1-\beta(\omega)) y$, where $\beta(\omega)=u(f(\omega))$. This mixing parameter exists, is unique and we have $h(\omega)\sim f(\omega)$. Recall that $f\in \Fudn$, so according to Proposition \ref{Proposition: characterization dependence neutrality for all notions} we have $\int u(f) \md q=0$ for all $q\in Q$ and hence $\int_{\Omega} \beta(\omega) \md q=\int u(f) \md q=0$. %
    Now, for an arbitrary linear $u^\prime \colon X \to \R$, we find
    \begin{align*}
        \int_{\Omega} u^\prime (h) \md q &= \int_{\Omega} u^\prime (h(\omega)) \md q(\omega)\\
        &= \int_{\Omega} u^\prime(\beta(\omega) x + (1-\beta(\omega))y) \md  q(\omega) \\
        &= \int_{\Omega} \left[\beta(\omega) u^\prime(x) + (1-\beta(\omega))u^\prime(y)\right] \md q(\omega) \\
        &= u^\prime(x)\underbrace{\int_{\Omega} \beta(\omega) \md q(\omega)}_{=0} + u^\prime(y) \underbrace{\int_{\Omega} (1-\beta(\omega)) \md q(\omega)}_{=\int 1 \md q - \int \beta(\omega) \md q(\omega)=0} = 0.
    \end{align*}
Thus, $h\in \FvNMdn$ as required. 
    
\end{proof}
    
\end{alemma}

We are now ready to prove Theorem \ref{theorem: dependence uncertainty}

\begin{proof}[Proof of Theorem \ref{theorem: dependence uncertainty}] 
    \emph{\eqref{item: representation result representation} $\Rightarrow$ \eqref{item: representation result axioms}:}
    Assume that $\succcurlyeq$ admits a dependence uncertainty representation. Clearly, Axiom \ref{axiom: non-trivial weak order} is satisfied. For the Archimedean axiom \ref{axiom: dn archimedean}, consider $g\in \mathcal{F}$, $f,h\in \FvNMdn$ and assume that $f\succ g\succ h$. Let ${\underline{p}}_g \in \arg\min_{p\in \mathcal{P}} \int u(g) \md p$. Since $f,h \in \FvNMdn$, this is also true of $\lambda f + (1-\lambda) h$. Therefore, $V(\lambda f + (1-\lambda)h) = \lambda V(f) + (1-\lambda) V(h)$ for any $\lambda \in (0,1)$, i.e., $V$ is linear on $\FvNMdn$. Since $V(f) > V(g) > V(h)$, we therefore can find desired scalars $\lambda, \kappa \in (0,1)$. %
    Axiom \ref{axiom: unboundedness} follows since $V$ is linear on $X \subseteq \FvNMdn$ with $V(x) = u(x)$ and $u$ is not bounded from above and below. %
    Axiom \ref{axiom: dn independence} follows directly from the above insight that $V$ is linear on $\FvNMdn$. %
    For Axiom \ref{axiom: dn dominance}, consider $f\in \mathcal{F}$ and $g\in \FvNMdn$ such that $f\geq g$, i.e., $u(f(\omega)) \geq u(g(\omega))$ for all $\omega \in \Omega$. Let $p^*\in \arg\min_{p\in \mathcal{P}} \int_\Omega u(f)\md p$. We have, by monotonicity of the integral, $V(f) \geq \int_\Omega u(f)\md p^* \geq \int_\Omega u(g)\md p^* = V(g)$ and thus $f\succcurlyeq g$. The ``$\leq$"-direction is analogous.
     
    \emph{\eqref{item: representation result axioms} $\Rightarrow$ \eqref{item: representation result representation}:}
      We will step by step construct the functional $V$, which will be of the dependence uncertainty type and represent $\succcurlyeq$. Since all constant acts are in $\FvNMdn$, Axiom \ref{axiom: dn independence} implies the full independence axiom on $X$. Further, Axiom \ref{axiom: dn archimedean} implies the Archimedean axiom on $X$. The axiom of the von Neumann-Morgenstern theorem are thus all satisfied on $X$, which implies the existence of a non-constant affine $u:X\rightarrow \mathbb{R}$, unique up to positive affine transformations, which represents the risk preferences reflected by $\succcurlyeq$. For $i\in \{1, \ldots, n\}$, consider the relation $\succcurlyeq_i$ on $\mathcal{F}_i$ defined by $f_i\succcurlyeq_i g_i \iff f_i\succcurlyeq g_i$ for all $f_i, g_i\in \mathcal{F}_i$. Since all acts in $\mathcal{F}_i$ are also in $\FvNMdn$, Axiom \ref{axiom: dn independence} implies the full independence axiom on $\mathcal{F}_i$. Furthermore, Axiom \ref{axiom: dn dominance} implies full monotonicity of $\succcurlyeq_i$ and Axiom \ref{axiom: dn archimedean} implies the standard Archimedean axiom on $\succcurlyeq_i$. Therefore, this sub-relation satisfies all the axioms of the Anscombe-Aumann theorem, implying the existence of a unique $p_i\in \Delta(\Omega_i)$ such that $\succcurlyeq_i$ is represented by the functional $V_i:\mathcal{F}_i \rightarrow \mathbb{R}$, $V_i(f_i) = \int_\Omega u_i(f_i) \md p_i$. Now take any $p^*\in \mathcal{P}:=\mathcal{P}(\Omega;p_1, \ldots, p_n)$. We have $V_i(f_i) = \int_\Omega u(f_i) \md p^*$ for all $f_i\in \mathcal{F}_i$ since $p^*$ has the marginal $p_i$ by the definition of $\mathcal{P}$. Define $V(f_i)=V_i(f_i)$. Consider an arbitrary $f\in \FvNMdn$. According to Proposition \ref{Proposition: characterization dependence neutrality for all notions}, $f$ is the mix of $\mathcal{F}_i$ acts, i.e., $f=\sum_{i=1}^n \alpha_i f_i$. Through standard arguments, Axiom \ref{axiom: dn independence} implies that $V$ is linear on $\FvNMdn$, thus $V(f)= \sum_{i=1}^n \alpha_i V(f_i)= \sum_{i=1}^n\int 
      \alpha_i u(f_i) \md p^* = \int u(f) \md p^*$. 
      
It remains to show that $\alpha(f)$ exists and is unique for all $f\in \mathcal{F} \setminus \FudnO(\Omega; p_1,\ldots,p_n)$. %
Consider $\overline f, \underline f\in \FvNMdn$ with $\overline{f} \geq f \geq \underline{f}$, such that $V(\underline f) = \min_{p\in \mathcal{P}} \int u(f) \md p$ and $V(\overline f) = \max_{p\in \mathcal{P}} \int u(f) \md p$. These acts can be constructed by combining the auxiliary Proposition \ref{AProposition: algorithm to find rectangular equivalent} and Lemma \ref{lemma: Fudn to FvNMdn}. Axiom \ref{axiom: dn dominance} implies $\overline{f} \succeq f \succeq \underline{f}$. We show that there exists a $\beta^* \in [0,1]$ such that $f\sim \beta^* \overline f + (1-\beta^*) \underline f$. Since $f\notin \FudnO(\Omega; p_1, \ldots, p_n)$ we know that $\overline f \succ \underline f$. Thus, for $0 \leq a<b \leq 1,$ we have $b \overline f+(1-b) \underline f \succ a \overline f+(1-a) \underline f$. This ensures that if $\beta^*$ exists, it is unique.

Since $f\in \mathcal{F} \setminus \FudnO(\Omega; p_1,\ldots,p_n)$, we have either $\underline{f} \prec f$ or $f \prec \overline{f}$ and thus uniquely $\beta^* = 0$ resp.\ $\beta^* = 1$ if $\underline{f} \sim f$ resp.\ $\overline{f} \sim f$. Define
$
    \beta^*=\sup \left\{\beta \in[0,1]: f \succeq \beta \overline f+(1-\beta) \underline f\right\}   
$. 
 Since $\beta=0$ is an element, the set is non-empty. By definition, if $1 \geq \beta>\beta^*$, then $f \prec \beta \overline f+(1-\beta) \underline f$. Moreover, just as for uniqueness above, if $0 \leq \beta<\beta^*$, then $f \succ \beta \overline f+(1-\beta) \underline f$. To see this, note that if $0 \leq \beta<\beta^*$, then there exists $\beta^{\prime}$ such that $0 \leq \beta<\beta^{\prime} \leq \beta^*$ and $f \succsim \beta^{\prime} \overline f+\left(1-\beta^{\prime}\right) \underline f$ by definition of $\beta^*$. Note that $\beta<\beta^{\prime}$ implies that $f \succsim \beta^{\prime} \overline f+\left(1-\beta^{\prime}\right) \underline f \succ \beta \overline f+(1-\beta) \underline f$.

First assume $\beta^* \overline f+\left(1-\beta^*\right) \underline f \succ f \succ \underline f$. By Axiom \ref{axiom: dn archimedean} there exists some $\lambda \in(0,1)$ such that $\lambda\left[\beta^* \overline f+\left(1-\beta^*\right) \underline f\right]+(1-\lambda) \underline f=\lambda \beta^* \overline f+(1- \left.\lambda \beta^*\right) \underline f \succ f$. But $\lambda \beta^*<\beta^*$, so by the previous argument $f \succ \lambda \beta^* \overline f+(1- \left.\lambda \beta^*\right) \underline f$, a contradiction.

Now assume $\overline f \succ f \succ \beta^* \overline f+\left(1-\beta^*\right) \underline f$. By Axiom \ref{axiom: dn archimedean}, there exists some $\mu \in(0,1)$ such that $f \succ \mu\left[\beta^* \overline f+\left(1-\beta^*\right) \underline f\right]+(1-\mu) \overline f= \left(1-\mu\left(1-\beta^*\right)\right) \overline f+\mu\left(1-\beta^*\right) \underline f$. Since $\left(1-\mu\left(1-\beta^*\right)\right)>\beta^*$, we have from above that $\left(1-\mu\left(1-\beta^*\right)\right) \overline f+\mu\left(1-\beta^*\right) \underline f \succ f$, a  contradiction.

This implies $f \sim \beta^* \overline f+\left(1-\beta^*\right) \underline f$. We have thus shown that we can choose 
\begin{equation*}
    V(f)=V\left(\beta^* \overline f+\left(1-\beta^*\right) \underline f\right) = \beta^* V(\overline f) + (1-\beta^*) V(\underline f),    
\end{equation*}

where the second equality follows from the fact that $V$ is linear on $\Fudn$. We thus have the required 

\begin{equation*}
    V(f)=\beta^* \max\limits_{p \in \mathcal{P}} \int u(f) \md p +\left(1-\beta^*\right) \min\limits_{p \in \mathcal{P}} \int u(f)\md p. 
\end{equation*}

So $\alpha(f)=1-\beta^*$ works and is uniquely determined.
\end{proof}

\subsubsection*{Proofs of Section \ref{section: MEU and smooth}}

\begin{proof}[Proof of Proposition \ref{prop: dependence uncertainty MEU and smooth}]
\emph{\eqref{item: MEU subset dependence uncertainty}} Assume that $\mathcal{C} \not\subseteq \mathcal{P}$. This implies the existence of some $\omega_i\in \Omega_i$ such that $\min\limits_{p\in \mathcal{C}} p([\omega_i])<\max\limits_{p\in \mathcal{C}} p([\omega_i])$, i.e., $\min\limits_{p\in \mathcal{C}} p([\omega_i])+\min\limits_{p\in \mathcal{C}} p([\omega_i]^c) <1$. We show that this violates Axiom \ref{axiom: dn independence}. %
Consider the acts $f_i= x_{[\omega_i]}z$ and $g_i=z_{[\omega_i]}y$ in $\mathcal{F}_i$, where $u(x)=\min\limits_{p\in \mathcal{C}} p([\omega_i]^c)$, $u(y)=\min\limits_{p\in \mathcal{C}} p([\omega_i])$ and $u(z)=0$. These consequences exist after an affine rescaling of $u$. By construction we have $f_i\sim g_i$, since $V(f_i) = \min\limits_{p\in \mathcal{C}} p([\omega_i]) \cdot \min\limits_{p\in \mathcal{C}} p([\omega_i]^c) =V(g_i)$. %
For $\lambda := \tfrac{\min\limits_{p\in \mathcal{C}} p([\omega_i])}{\min\limits_{p\in \mathcal{C}} p([\omega_i]) + \min\limits_{p\in \mathcal{C}} p([\omega_i]^c)}$, consider the act $h_i=\lambda f_i + (1-\lambda) g_i$. %
The act $h_i \in \mathcal{F}_i$ has constant utility $\tfrac{\min\limits_{p\in \mathcal{C}} p([\omega_i]) \cdot \min\limits_{p\in \mathcal{C}} p([\omega_i]^c)}{\min\limits_{p\in \mathcal{C}} p([\omega_i]) + \min\limits_{p\in \mathcal{C}} p([\omega_i]^c)}$. Note that the denominator is strictly smaller than 1 by assumption, thus the fraction is strictly bigger than $\min\limits_{p\in \mathcal{C}} p([\omega_i]) \cdot \min\limits_{p\in \mathcal{C}} p([\omega_i]^c)$. We thus have $f_i \sim g_i$, but $h\succ g_i$, contradicting the Axiom \ref{axiom: dn independence}.\\

\emph{\eqref{item: smooth subset dependence uncertainty}} Assume by means of contradiction that $\supp(\mu) \not\subseteq \mathcal{P}$. Then there is an $\omega_i$ with
$
    \mu(p\in \Delta(\Omega) | p([\omega_i]) = p_i(\omega_i))<1
$. %
We now show that this contradicts Axiom \ref{axiom: dn independence}. %
The non-linearity of $\phi$ implies that the second derivative $\phi^{\prime\prime}$ is not constant $0$. Since $\phi^{\prime\prime}$ is continuous, there exists an interval, by suitable transformations wlog $[0,1] \subseteq u(X)$, on which $\phi^{\prime\prime}$ is strictly positive or negative. Thus, $\phi$ is strictly concave or strictly convex on $[0,1]$. Assume the latter (the concavity case is analogous). %
Let $z^1,z^0 \in X$ with $u(z^0) = 0, u(z^1) = 1$. Consider $f_i := {z^1}_{[\omega_i]}z^0$. Then, using Jensen's inequality and $\mu(p\in \Delta(\Omega) | p([\omega_i]) = p_i(\omega_i))<1$, we find $V(f_i) = \int_{\Delta(\Omega)} \phi (p([\omega_i])) \md \mu(p) > \phi(\int_{\Delta(\Omega)} p([\omega_i]) \md \mu(p))$. Let $x \in X$ be such that $m:= u(x) = \int_{\Delta(\Omega)} p([\omega_i]) \md \mu(p)$ and set $g_i  \equiv x$. Thus, $f_i \succ g_i$. Now, let $\alpha \in (0,1)$ be small such that $\tfrac{m-\alpha}{1-\alpha}, \tfrac{m}{1-\alpha} \in (0,1)$. Let $x',x'' \in X$ such that $u(x') = \tfrac{m-\alpha}{1-\alpha}, u(x'') =  \tfrac{m}{1-\alpha}$. Set $h_i :=  x'_{[\omega_i]} x''$. Using Jensen's inequality again, we find $V(\alpha g_i + (1-\alpha)h_i) = \int_{\Delta(\Omega)} \phi(\alpha m + m - \alpha p([\omega_i]))\md \mu(p) > \phi(\alpha m + m - \alpha m) = \phi(m) = V(\alpha f_i + (1-\alpha)h_i)$. Consequently, $\alpha f_i + (1-\alpha)h_i \prec \alpha g_i + (1-\alpha)h_i$, violating Axiom \ref{axiom: dn independence}.

\emph{\eqref{item: variational characterization dependence uncertainty}} %
Assume that $\succcurlyeq$ is variational and satisfies axioms \ref{axiom: non-trivial weak order} - \ref{axiom: dn dominance}. Fix $p\notin\mathcal P$. There exist $i\in \{1, \ldots, n\}$ and $\omega_i\in\Omega_i$ such that $p([\omega_i]) < p_i(\omega_i)$. For every $t\in\mathbb R$, choose $f_t\in\mathcal F_i$ with utility profile $u(f_t)=\one_{[\omega_i]}$. Such acts exist because $u(X)=\mathbb R$, which follows by Unboundedness. We have $V(f_t)=t p_i(\omega_i)$. Hence by definition of $c^*$
\begin{equation*}
    c^*(p)\geq V(f_t)-\int_\Omega u(f_t)\md p=t\bigl(p_i(\omega_i)-p([\omega_i])\bigr).
\end{equation*}
Letting $t$ tend to infinity gives $c^*(p)=\infty$. Finally, grounding and lower semicontinuity on the compact simplex imply that $c^*$ attains zero, which must happen for some element in $\mathcal{P}$.
\end{proof}

\subsubsection*{Proofs of Section \ref{Section: dependence neglect}}
\begin{proof}[Proof of Theorem \ref{theorem: Full Independence}]
    \emph{\eqref{item: Theorem full independence SEU pind} $\Rightarrow$ \eqref{item: Theorem full independence subspace separation}:} Let $\mathcal{C} = \{\pind\}$. Then we have for all $i \in \{1, \ldots, n\}$, $f_i \in \mathcal{F}_i$, $[E_{-i}]$ non-null w.r.t.\ $\succeq$ and $x \in X$
    \begin{align*}
        V({f_i}_{E_{-i}}x) =&  \sum_{\omega_i \in \Omega_i} u(f_i(\omega_i)) \cdot p_i(\omega_i) \cdot \pind \res_{\Omega_{-i}} (E_{-i}) + u(x) \cdot (1-p \res_{\Omega_{-i}}(E_{-i}))\\
        =& V(f_i) \cdot \pind\res_{\Omega_{-i}}(E_{-i}) + u(x) \cdot (1-\pind \res_{\Omega_{-i}}(E_{-i})).
    \end{align*}
    Since $\pind\res_{\Omega_{-i}}(E_{-i}) = \pind([E_{-i}]) \neq 0$, we see that Axiom \ref{axiom:subspace separation} is fulfilled.

    \emph{\eqref{item: Theorem full independence subspace separation} $\Rightarrow$ \eqref{item: Theorem full independence SEU pind}:} Let $i \in \{1, \ldots, n\}$, $x \succ y, {x'} \succ y$ with $u(x) = 1, u(y) = 0$, $[E_{-i}]$ be non-null, $E_i \subseteq \Omega_i$ arbitrary and define ${x'}$ such that we get the indifference $f^1 := x_{[E_i \times E_{-i}]}y \sim {x'}_{[\Omega_i \times E_{-i}]}y=:g^1$. From Axiom \ref{axiom:subspace separation}, we thus also get the indifference $f^2 := x_{[E_i \times \Omega_{-i}]}y \sim {x'}_{[\Omega_i \times \Omega_{-i}]}y=:g^2$. The last indifference is equivalent to $u({x'}) = V(g^2)= V(f^2) = \min_{p \in \mathcal{C}} p([E_i]) = p_i(E_i)$, since $p \res_{\Omega_i} = p_i$ for all $p \in \mathcal{C} \subseteq \mathcal{P}(\Omega; p_1, \ldots, p_n)$. Plugging this into the first indifference, we obtain
    \begin{equation*}
        \min_{p \in \mathcal{C}} p(E_i \times E_{-i}) = V(f_1) = V(g^1) = u({x'}) \min_{p \in \mathcal{C}} p([E_{-i}]) = p_i(E_i) \min_{p_{-i} \in \mathcal{C}\res_{\Omega_{-i}}} p_{-i}(E_{-i}).
    \end{equation*}
    The above argument shows that we can cut off $p_i$ as an independent component when calculating worst-case expectations with respect to events of the form $E_i \times E_{-i}$.
    Applying this argument successively to the other indices (with keeping $\Omega_i$ in the $i$-component of a non-null new event $E_{-j} = \Omega_i \times E_{-\{j,i\}}$) we eventually find
    \begin{equation*}
        \min_{p \in \mathcal{C}} p(E_1 \times \ldots \times E_n) = \prod_{i=1}^n p_i(E_i) = \pind(E_1 \times \ldots \times E_n) \label{eq: minimium related to pind}
    \end{equation*}
    for all $E_i \subseteq \Omega_i$. If there was $\pind \neq p^* \in \mathcal{P}$, then we would find an $\omega \in \Omega$ with $\min_{p \in \mathcal{C}}p(\omega) \leq p^*(\omega) < \pind (\omega)$. Consequently, $\pind \in \mathcal{C}$ and $\mathcal{C} = \{ \pind \}$.
\end{proof}

\subsubsection*{Proofs of Section \ref{section: comparative dependence uncertainty and dependence premia}}

\begin{proof}[Proof of Proposition \ref{Proposition: Comparative dependence aversion}]
\emph{\eqref{item: comparative dependence def} $\Rightarrow$ \eqref{item: comparative dependence functional}}: Assume that $\succcurlyeq$ is more dependence averse than $\succcurlyeq^\prime$. This implies that the preference relations agree on $\FvNMdn$. They thus agree on constant acts, thus $u^\prime$ is a positive affine transformation of $u$. Without loss, assume that $u=u'$. They also agree on $\mathcal{F}_i$ for all $i\in \{1, \ldots, n\}$. Thus they induce the same marginals on the subspaces, so $p_i=p_i^\prime$ for all $i\in \{1, \ldots, n\}$. Now consider some $f\in \mathcal{F}\backslash \Fudn$. Let $g\in \FvNMdn$ be such that $f\sim g$, e.g., a certainty equivalent. %
Since $\succcurlyeq$ is more dependence averse than $\succcurlyeq^\prime$ it follows that $f\succcurlyeq^\prime g$. From the formula of dependence uncertainty (see \eqref{equation: DU model intro}), it follows that $\alpha^\prime(f) \leq \alpha(f)$.

\noindent%
\emph{\eqref{item: comparative dependence functional}
$\Rightarrow$\eqref{item: comparative dependence def}}: Assume that $u^\prime=u$ and $p_i = p_i^\prime$. Consider an arbitrary $f\in \mathcal{F} \setminus \Fudn$, $g\in \FvNMdn$ and assume $f\succcurlyeq g$. Since $g \in \FvNMdn$, we have $V(g) = V'(g)$ for the respective representation functionals. Finally, from $\alpha(f)\geq \alpha^\prime(f)$, we find $V'(f) \geq V(f) \geq V(g) = V'(g)$.

\noindent%
Let now $X = \R$, $u=u' \colon \R \to \R$ be strictly increasing and continuous and $p_i = p_i'$ for all $i$. Suppose first that $\DP(f) \geq \DP'(f)$ for all $f \in \mathcal{F}$. Let $f \succcurlyeq g$ with $g \in \FvNMdn$. From the definition of the dependence premium, expression \eqref{eq: dependence premium definition}, we obtain $V'(f) \geq V(f) \geq V(g) = V'(g)$ and thus \eqref{item: comparative dependence def}. For the other direction, suppose \eqref{item: comparative dependence functional}, i.e., $\alpha(f) \geq \alpha'(f)$ for all $f \in \mathcal{F}$. Therefore, $\int_{\Omega} u(f-\DP(f) \md \pind = V(f) \leq V'(f) = \int_{\Omega} u(f-\DP'(f)) \md \pind$ and thus $\DP'(f) \leq \DP(f)$.
\end{proof}
\begin{proof}[Proof of Corollary \ref{corollary: comparative DP meu and smooth}]

By Proposition \ref{Proposition: Comparative dependence aversion},
comparability in dependence aversion requires positive affinely related
utilities and identical marginal beliefs. Whenever utility indices are
positive affinely related, we express parameter comparisons in common
utility units. Recall that every profile $a \in \R^{\Omega}$ can be replicated by the utility profile $u(f)$ of an act $f$ by Axiom \ref{axiom: unboundedness}. We use the scalar product $\langle \cdot , p \rangle$ to compactly write integration over $\Omega$ with respect to a probability $p$. In order to leverage Proposition \ref{Proposition: Comparative dependence aversion}, we re-write the dependence uncertainty representations in the form of \eqref{equation: DU model intro}, comparing the $\alpha$ values.\\
\emph{\eqref{cor comp DP MEU}:} If $\mathcal{C} \supseteq \mathcal{C}^\prime$, it is straightforward to see that $\alpha^\prime(f) \leq \alpha(f)$ in the representation. Conversely, if $p^\prime \in \mathcal{C}^\prime \setminus \mathcal{C}$, apply the hyperplane separation theorem to obtain an act $f$ with utility profile $u(f)$ such that $\langle u(f), p^\prime \rangle < \langle u(f), p\rangle$ for all $p \in \mathcal{C}$. Since $\min_{\hat{p} \in \mathcal{C}^\prime} \langle u(f), \hat{p} \rangle \leq \langle u(f), p^\prime \rangle < \langle u(f), p \rangle$ for all $p \in \mathcal{C}$, it follows that $\alpha^\prime(f) > \alpha(f)$. The statement about dependence aversion now follows directly from Proposition \ref{Proposition: Comparative dependence aversion}.\\
\emph{\eqref{cor comp DP smooth}:} Set $\tilde{V} (f) := \phi^{-1}(V(f))$ and likewise $\tilde{V'}$ to obtain a representation that is bounded from above and below by the maximum and minimum expectation of $u(f)$ over $\mathcal{P}$. Now, if $h := \phi \circ (\phi')^{-1}$ is concave, we have $\tilde{V}(f) \leq \tilde{V'}(f)$ for any act $f$ by applying Jensen's inequality. Consequently, $\alpha(f) \geq \alpha^\prime(f)$ for all $f \in \mathcal{F}$. Conversely, if $h$ is not concave, there is $t_0 \in \R$ such that $h''(t_0)>0$, i.e., $h$ is locally strictly convex around $t_0$ by our smoothness assumptions. Let $t_1,t_2$ be to distinct points in this convex region. If $\mu$ is not degenerate, there is a measurable set $A \subseteq \Delta(\Omega)$ with $0 < \mu(A) < 1$. Now choose $f$ such that $u(f)$ is equal to $t_1$ on $A$ and equal to $t_2$ otherwise. Applying Jensen's inequality, we now get $\tilde{V}(f) > \tilde{V'}(f)$ and thus $\alpha(f) < \alpha^\prime(f)$. Now apply Proposition \ref{Proposition: Comparative dependence aversion}. If $\mu = \delta_p$ for a single measure $p$, then $\tilde{V}(f) = \int_{\Omega}u(f) \md p$ irrespective of $\phi$, i.e., all smooth preferences with affinely related $u$ and average prior $p$ are identical.\\
\emph{\eqref{cor comp DP VP}:} Since the minimal cost functions are grounded, $V(f),V'(f)$ are bounded from above and below by $\max_{p \in \mathcal{P}} \int_{\Omega} u(f) \md p$ resp.\ $\min_{p \in \mathcal{P}} \int_{\Omega} u(f) \md p$. If now $c^* \leq {c^*}^\prime$ then $V(f) \leq V'(f)$ and thus $\alpha(f) \geq \alpha'(f)$ for all $f \in \mathcal{F}$. Conversely, if $\alpha(f) \geq \alpha'(f)$, then $V(f) \leq V'(f)$. By minimality of the cost functions, $c^*(p) = \sup_{f \in \mathcal{F}} \{ V(f) - \int_{\Omega} u(f) \md p \} \leq \sup_{f \in \mathcal{F}} \{ V'(f) - \int_{\Omega} u(f) \md p \} = {c^*}^\prime(p)$.
\end{proof}

\begin{proof}[Proof of Corollary \ref{Corollary: Absolute dependence aversion}]
\emph{\eqref{item: Cor ADA MEU}:} Follows readily from Corollary \ref{corollary: comparative DP meu and smooth}\eqref{cor comp DP MEU}.\\%
\emph{\eqref{item: Cor ADA smooth}:} If $\mu$ is non-degenerate, $\succcurlyeq$ is more dependence averse than the SEU representation $\succcurlyeq_{\bar{p_{\mu}}}$ with utility $u$, and prior $\bar{p}_{\mu}$ if and only if $\phi$ is concave by Proposition \ref{Proposition: Comparative dependence aversion}, choosing the identity function as ambiguity index. To see that $\bar{p}_{\mu} = \pind$ if $\succcurlyeq$ is more dependence averse than $\succcurlyeq_{\pind}$, %
let $f_{a,t,v} \in \mathcal{F}$ have utility profile $u(f) = a \cdot \boldsymbol{1} + t \cdot v$ for given $a,t \in \R$ and $v \in \R^{\Omega}$. Since $\succcurlyeq$ is more dependence averse than $\succcurlyeq_{\pind}$, $H_{a,v}(t) := \int_{\Delta(\Omega)} \phi(a + t \langle v, p \rangle) \md \mu(p) - \phi(a + t \langle v, \pind \rangle) \leq 0$. Consequently, $H_{a,v}(0) = 0$, $t=0$ is a global maximum for any fixed $a,v$. Applying the Leibniz rule, yields $0 = H'_{a,v}(t)\res_{t=0} = \int_{\Delta(\Omega)}\phi'(a) \langle v,p \rangle \md \mu(p) - \phi'(a) \langle v, \pind \rangle = \phi'(a) \langle v, \bar{p}_{\mu} - \pind \rangle$. Since $\phi$ is strictly increasing, a choice of $a$ with $\phi'(a) >0$ reveals $\bar{p}_{\mu} = \pind$. This argument also holds if $\mu$ is degenerate.\\
\emph{\eqref{item: Cor ADA VP}:} Regard $\succcurlyeq_{\pind}$ as a variational preference with minimal cost function $c^*_{\pind} (p) = 0$ if $p = \pind$ and $c^*_{\pind} (p) = \infty$ otherwise. Then, $\succcurlyeq$ is more dependence averse than $\succcurlyeq_{\pind}$ if and only if $c^* \leq c^*_{\pind}$, which is equivalent to $c^*(\pind) = 0$.
\end{proof}

\subsubsection*{Proofs of Section \ref{Section: dependence-invariance}}
\begin{proof}[Proof of Proposition \ref{proposition: dependence equivalence characterization}]
We first prove the second, more general statement.\\
\emph{\eqref{item: bowtie mixture characterization}:} %
Assume $f \bowtie g$, i.e., $\int_{\Omega} u(f) \md q \gtreqqless 0$ if and only if $\int_{\Omega} u(g) \md q \gtreqqless 0$ for all affine $u$ and $q \in Q$. Consequently, there is $\kappa(u,q)>0$ for each affine $u$ and $q \in Q$ such that $\int_{\Omega} u(f) \md q = \kappa(u, q) \int_{\Omega} u(g) \md q$. We claim that $\kappa >0$ can be chosen as a constant. First, we claim that $\kappa(u,q)$ is constant in $q$ for any fixed $u$. It suffices to prove that $\kappa(u, q)$ has to be constant on $A := \{ q \in Q \mid \int_{\Omega} u(f) \md q \neq 0 \} $ (If the set is empty, simply choose $\kappa = 1$). To this end, let $q,q' \in A$ and $\kappa := \kappa(u,q), \kappa' := \kappa(u,q')$. By possibly switching to $-q$ and/or $-q'$, we may assume $\int_{\Omega} u(f) \md q >0$ and $\int_{\Omega} u(f) \md q' >0$. Then, there is $\lambda >0$ such that $\int_{\Omega} u(f) \md (q-\lambda q') = 0$. From this, we 1.\ get that also $\int_{\Omega} u(g) \md (q-\lambda q') = 0$ as $q-\lambda q' \in Q$ and 2.\ that $0 = \kappa \int_{\Omega} u(g) \md q - \lambda \kappa' \int_{\Omega} u(g) \md q'$. Together, these two observations result in $\kappa = \kappa'$. Applying the analogous argument on $B := \{ u \text{ affine} \mid \int_{\Omega} u(f) \md q \neq 0 \}$ for any fixed $q \in Q$, we see that $\kappa >0$ can be chosen constant with $\int_{\Omega} u(f) \md q = \kappa \int_{\Omega} u(g) \md q$ for all affine $u$ and $q \in Q$.\\%
Define now $h(\omega)(z) := f(\omega)(z) - \kappa g(\omega)(z) \in \R$. As in the proof of Proposition \ref{Proposition: characterization dependence neutrality for all notions}, define $g_i(\omega)(z) := h(\omega_i, \omega_{-i}^z)(z) - c^z$ with $\omega^z \in \argmin_{\omega} h(\omega)(z)$ and $c^z := h(\omega^z)(z)$ for each $i \in \{1 \ldots, n\}$ and $z$ for all $z$ with $h(\omega)(z) \neq 0$ for some $\omega$. Similar to before, we observe that $f(\omega)(z) - \kappa g(\omega)(z) = h(\omega)(z) = c^z + \sum_{i=1}^ng_i(\omega)(z)$ and that $\omega \mapsto \sum_{z} g_i(\omega)(z)$ is a constant $\hat{\gamma}_i \geq 0$. Set $f_i(\omega)(z) := \hat{\gamma}_i^{-1} \sum_{z} g_i(\omega)(z) \in \FvNMdn$ whenever possible, i.e., $\hat{\gamma}_i >0$. Set $x^+(z') := \eta'^{-1} \cdot \sum_{z \mid c^z \geq 0}c^z \one_z(z') \in X$, where $\eta' := \sum_{z \mid c^z \geq 0} c^z$, and $x^-(z') = \eta^{-1} \cdot \sum_{z \mid c^z < 0}- c^z \one_z(z') \in X$, where $\eta := \sum_{z \mid c^z < 0} -c^z$. Let $\beta := (1 + \eta)^{-1}$ and $\beta' := \kappa (\kappa+\sum_{i=1}^n \hat{\gamma}_i + \eta')^{-1}$. Then, $\beta f + (1-\beta) x^-$ is equal to\\
$\beta' g + (1-\beta')\left[ (\sum_{j=1}^n \hat{\gamma}_j)/(\sum_{j=1}^n \hat{\gamma}_j + \eta')\sum_{i=1}^n\hat{\gamma}_i f_i + \eta'/(\sum_{j=1}^n \hat{\gamma}_j + \eta') x^+\right]$ as desired.\\
Conversely, let now $\beta,\beta'\in (0,1]$ and $h, h^\prime\in \FvNMdn$ be such that $\beta f + (1-\beta) h=\beta' g +(1-\beta') h^\prime$. Consider an arbitrary affine $u \colon X \to \mathbb{R}$ and an arbitrary $q\in Q$. Since $u$ is affine we have $\beta u(f) = \beta' u(g) + (1-\beta') u(h^\prime) - (1-\beta) u(h)$. Then, for any $q \in Q$, $\int_{\Omega} u(f) \md q = \tfrac{\beta'}{\beta} \int\limits u(g) \md q$ since $h,h' \in \FvNMdn$, i.e., $f \bowtie g$.\\
\emph{\eqref{item: asymp mixture characterization}:} %
This follows easily from the above proof, noting in particular that $f \asymp g$ implies $\kappa = 1$ and $c^z \leq 0$.
\end{proof}

\begin{proof}[Proof of Theorem \ref{theorem: dependence-invariant DU}]
The idea is that Axioms \ref{axiom: DN translation independence} and
\ref{axiom: DN affine independence} mimick the role that (Weak) Certainty Independence, by replacing constants with dependence-neutral acts. In particular, \ref{axiom: DN translation independence} leads to additivity of $V$ with respect to dependence-neutral utility profiles, whereas
\ref{axiom: DN affine independence} additionally yields positive homogeneity.\\
Recall that $u(X) = \R$ by Axiom \ref{axiom: unboundedness}. In particular, for any utility profile $a \in \R^\Omega$ there is $f \in \mathcal{F}$ with $u(f) = a$. By Axiom \ref{axiom: statewise consequentialism}, an act's evaluation hinges only on its utility profile. Thus, we can define a function $I \colon \R^{\Omega} \to \R$ through $V(f) = I(u(f))$. Let $\boldsymbol 1 \in \R^{\Omega}$ be the vector consisting of a constant utility level $1$. It follows that $I(r \boldsymbol 1) = r$ for all $r \in \R$. If $c \in u(\FvNMdn)$, i.e., $c=u(h)$ for some $h \in \FvNMdn$, then $I(c) = V(h) = \int_{\Omega} u(h) \md \pind$. In particular, if $c,c' \in u(\FvNMdn)$, then also $\tilde{c} := c+c' \in u(\FvNMdn)$ and thus $I(c+c') = I(\tilde{c}) = \int_{\Omega} \tilde{c}\md \pind = \int_{\Omega} c \md \pind + \int_{\Omega} c' \md \pind = I(c) + I(c')$. We first show sufficiency of the axioms.\\
Assume now Axiom \ref{axiom: DN translation independence} holds and let $a, b \in \R^\Omega$ and $c \in u(\FvNMdn)$. There are thus acts $f,g \in \mathcal{F}$, $h \in \FvNMdn$ such that $2a = u(f)$, $2b = u(g)$ and $2c = u(h)$. Then, $I(a+c) = V(\tfrac{1}{2}f + \tfrac{1}{2}h)$ and $I(b+c) = V(\tfrac{1}{2}g + \tfrac{1}{2}h)$. Applying Axiom \ref{axiom: DN translation independence} with $h' \in \FvNMdn$ with $u(h')= 0  \cdot \boldsymbol{1}$ yields $I(a+c) \geq I(b+c)$ if and only if $I(a) \geq I(b)$. Now, choosing $b = I(a) \boldsymbol1$, we have $I(a) = I(b)$. Consequently, using that $I(a) \boldsymbol{1}, c \in u(\FvNMdn)$, we find $I(a+c) = I(I(a) \boldsymbol{1} + c) = I(a) + I(c)$. If now $f \asymp g$, we have by Proposition \ref{proposition: dependence equivalence characterization} $\beta f + (1-\beta)h = \beta g + (1-\beta) h'$ and thus $u(f) = u(g) + \tfrac{1-\beta}{\beta} u(h'-h) =: u(g) + c$. Consequently, $V(f) = I(u(f)) = I(u(g)) + I(c) = V(g) + I(c)$. Note also that $\int_{\Omega}u(f) \md p = \int_{\Omega} u(g) \md p + I(c)$ for all $p \in \mathcal{P}$ and thus $\min_{p \in \mathcal{P}} \int_{\Omega} u(f) \md p = \min_{p \in \mathcal{P}} \int_{\Omega} u(g) \md p + I(c)$ and the same for the maximum over $p \in \mathcal{P}$. Combining these observations and unfolding $V$, yields $0 = (\alpha(f) - \alpha(g)) \cdot (\max_{p \in \mathcal{P}} \int_{\Omega} u(g) \md p - \min_{p \in \mathcal{P}} \int_{\Omega} u(g) \md p)$. Hence, if $g$ (and thus $f$) is not dependence-neutral, we obtain $\alpha(f) = \alpha(g)$. For dependence-neutral acts, any fixed choice $\alpha_0 \in [0,1]$ is feasible.\\
Assume now that Axiom \ref{axiom: DN translation independence}, and thus also \ref{axiom: DN affine independence}, holds. Consider $a \in \R^\Omega$ and $b = I(a) \boldsymbol{1}$ and acts $f,g,h$ with $a= u(f), b = u(g), c:=u(h) = 0 \cdot \boldsymbol{1}$. Since $V(f) = V(g)$, we find $V(\lambda f + (1-\lambda) h) = V(\lambda g + (1-\lambda)h)$ for any $\lambda \in (0,1)$ by applying Axiom \ref{axiom: DN translation independence}. Rewriting the equality and using the additivity w.r.t.\ $c$ from above yields $I(\lambda a) + I((1-\lambda)c) = I(\lambda I(a) \boldsymbol{1}) + I((1-\lambda)c)$, and thus $I(\lambda a) = \lambda I(a)$. If now $f \bowtie g$, then $u(f) = \tfrac{\beta'}{\beta} u(g) + u(\tilde{h})$ for some $\tilde{h} \in \FvNMdn$ by Proposition \ref{proposition: dependence equivalence characterization}. Consequently, $V(f) = I(u(f)) = I( \tfrac{\beta'}{\beta}u(g) + u(\tilde{h})) = \tfrac{\beta'}{\beta} V(g) + I(c)$ with $c = u(\tilde{h})$. Note also that $\int_{\Omega}u(f) \md p = \tfrac{\beta'}{\beta} \int_{\Omega} u(g) \md p + I(c)$ for all $p \in \mathcal{P}$. Combining these observations yields $(\alpha(f) - \alpha(g)) (\max_{p \in \mathcal{P}} \int_{\Omega} u(g) \md p - \min_{p \in \mathcal{P}} \int_{\Omega} u(g) \md p)$. Thus, if $f,g \notin \FvNMdn$, $\alpha(f) = \alpha(g)$. For dependence-neutral acts, any fixed choice $\alpha_0 \in [0,1]$ is feasible.\\
For necessity of the axioms, observe first that $V(\lambda f + (1-\lambda) h) = (1-\lambda) I(u(h)) + \lambda [ \alpha(\lambda f + (1-\lambda) h) \min_{p \in \mathcal{P}} \int_{\Omega} u(f) \md p + (1-\alpha(\lambda f + (1-\lambda)h)) \max_{p \in \mathcal{P}} \int_{\Omega} u(f) \md p]$ for $f \in \mathcal{F}$, $h \in \FvNMdn$. Note that $\lambda f + (1-\lambda) h \asymp \lambda f + (1-\lambda) h'$ for any $h' \in \FvNMdn$. If thus, $\alpha$ coincides on $\asymp$-equivalence classes, Axiom \ref{axiom: DN affine independence} follows. Likewise, note that $f \bowtie \lambda f + (1-\lambda) h$. Thus, if $\alpha$ coincides on $\bowtie$-equivalence classes, Axiom \ref{axiom: DN translation independence} follows.
\end{proof}

\subsubsection*{Proofs of Section \ref{Section: Math properties of P}}
\begin{proof}[Proof of Proposition \ref{Proposition: Dimension of P}]
    We prove $\dim(Q) = \prod_{i=1}^n \#\Omega_i - 1 -\sum_{i=1}^n (\Omega_i-1)$. The dimension of $\mathcal{P}$ follows from considering $Q^* = Q(\supp(p_1) \times \ldots \times \supp(p_n))$.
    
    Let $d(\omega, \omega') := \# \{ i \in \{1, \ldots, n\} \mid \omega_i \neq \omega'_i \}$ denote the \emph{Hamming distance}. In the following, fix any $\overline{\omega} \in \Omega$. Observe now that
    $
        \# \left.\left\{ \omega \in \Omega \right| d(\omega, \overline{\omega}) \geq 2 \right\}
        = \# \Omega - \# \left.\left\{ \omega \in \Omega \right| d(\omega, \overline{\omega}) \leq 1 \right\} =  \prod_{i=1}^n \#\Omega_i - 1 -\sum_{i=1}^n (\Omega_i-1)
    $. 
    For any $\omega$ with $d(\omega, \overline{\omega}) \geq 2$, we can pick any two distinct indices $i,j$ for which $\omega, \overline{\omega}$ disagree and consider $\omega^* := \omega$ and $\omega^{\dagger} := (\overline{\omega}_i, \overline{\omega}_j, \omega_{-ij})$. Note that scalar multiples of the rectangular operation defined by $q(\omega^*) = q(\omega^{\dagger}) = 1$, $q(\omega_i^*, \omega_{-i}^\dagger) = q(\omega_i^{\dagger}, \omega_{-i}^*) = -1$ allow us to eject all (positive or negative) mass from $\omega$ to elements in $\Omega$ with strictly smaller Hamming distance than $d(\omega, \overline{\omega})$. Collect all these rectangular operations for each element $\omega$ with $d(\omega,\overline{\omega}) \geq 2$ and call its set $\mathcal{B}$. We claim that $\mathcal{B}$ is a basis of $Q$:\\
    First, observe that $\mathcal{B}$ generates $Q$. Consider an arbitrary dependence structure $\overline{q} \in Q$. Iteratively eject all mass from those $\omega$ with $d(\omega,\overline{\omega}) \geq 2$ by applying scaled operations from $\mathcal{B}$, starting with the elements of maximal distance. The resulting $\overline{q} \in Q$ can only have non-zero mass on $\{ \omega \in \Omega \mid d(\omega, \overline{\omega}) \leq 1\}$. We claim $\overline{q} = 0$. To this end, take any $\omega$ with $d(\omega, \overline{\omega}) = 1$ and write $\omega = (\omega_i, \overline{\omega}_{-i})$. Since $\overline{q} \in Q$ and $q(\omega') = 0$ if $d(\omega', \overline{\omega}) \geq 2$, we find
    $
        0 = q([\omega_i]) = \sum_{\omega_{-i} \in \Omega_{-i}} q(\omega_i, \omega_{-i}) = q(\omega_i, \overline{\omega}_{-i}) = q(\omega)
    $. %
    Eventually, also $0 = q([\overline{\omega}_i]) = \sum_{\omega_{-i}} q(\overline{\omega}_i, \omega_{-i}) = q(\overline{\omega})$, thus, $\overline{q} = 0$. Applying the above operations from $\mathcal{B}$ backwards, we have therefore shown that $\mathcal{B}$ spans $Q$.\\
    Second, we cannot generate $Q$ from $\mathcal{B}$ if leaving out any of its rectangular operations w.r.t.\ an $\omega'$ with $d(\omega',\overline{\omega}) \geq 2$. To see this, note that whenever choosing the indices $i,j$ for an $\omega$ with $d(\omega,\overline{\omega}) > d(\omega', \overline{\omega})$, we can do so in a way that the respective rectangular operation w.r.t.\ $\omega$ fulfills $q(\omega') = 0$. Consequently, there no longer is a way to shift away mass from $\omega'$.
\end{proof}

\begin{proof}[Proof of Theorem \ref{Theorem: extreme points via mutual information}]
   Our proof of the equivalences is by ring-inference. As strictly decreasing transformation of a function have the same local maxima, it is sufficient to prove the statements for $f$- and Bregman divergences with infinitely decreasing slope in $0$. We consider only the case of an $f$-divergence, as the proof for a Bregman divergence is qualitatively the same.\\
    \emph{\eqref{item: p extreme} $\Rightarrow$ \eqref{item: p maximally zero}:} If $p$ is not maximally zero, there is $p' \in \mathcal{P}$, $p\neq p'$, which vanishes in all $\omega$ in which $p$ vanishes (and possibly more). In other words $p^{-1}(0) \subseteq p'^{-1}(0)$. Note that $x := p'-p \in Q \setminus \{ 0 \}$ and that $x(\omega) = 0$ whenever $p(\omega) = 0$. We can thus scale down $x$ by a scaling factor $\kappa >0$ such that $p+\kappa x, p - \kappa x \in \mathcal{P}$ (note that $\kappa x \in Q$). Consequently, $p$ is not an extreme point.\\
    \emph{\eqref{item: p maximally zero} $\Rightarrow$ \eqref{item: p local maximizer of mutual information}:} Let $p$ be maximally zero. We show that $\Divergence{f}{.}{\pind}$ is strictly increasing locally if moving towards $p$ from any direction. To this end, fix $p' \in \mathcal{P}$ with $p \neq p'$ and consider the convex combination $\lambda p + (1-\lambda)p'$ for $\lambda \in (0,1)$. After removing states with $p(\omega) = p'(\omega)$, we have
    \begin{align}
        &\frac{\partial}{\partial \lambda} \Divergence{f}{\lambda p + (1-\lambda) p'}{\pind}\nonumber\\
        =& \sum_{\substack{{\omega \in \Omega}\$p(\omega),p'(\omega)) \neq (0,0)}} (p(\omega) - p'(\omega)) \cdot \pind(\omega) \cdot f'\left( \frac{\lambda p(\omega) + (1-\lambda) p'(\omega)}{\pind(\omega)} \right).\label{eq: derivative mutual info convex combination}
    \end{align}
    We now claim that $\lim_{\lambda \to 1}\frac{\partial}{\partial \lambda} \Divergence{f}{\lambda p + (1-\lambda)p'}{\pind} = + \infty$, showing that $\Divergence{f}{.}{\pind}$ strictly increases locally when moving towards $p$. To see this, we make a case distinction w.r.t.\ $\omega$ for each summand. If $p(\omega) \neq 0$, the respective summand in expression \eqref{eq: derivative mutual info convex combination} has a finite limit for $\lambda \to 1$. If $p(\omega) = 0$ (and $p'(\omega) \neq 0$), the limit of the corresponding summand in \eqref{eq: derivative mutual info convex combination} is
    \begin{align}
        \lim_{\lambda \to 1} -p'(\omega) \cdot \pind(\omega) \cdot f'\left( \frac{(1-\lambda) p'(\omega)}{\pind(\omega)} \right) = +\infty,
    \end{align}
    as $\lim_{t \to 0} f'(t) = -\infty$. %
    Since $p$ is maximally zero, there indeed exists at least one $\omega$ with $p(\omega) = 0 \neq p'(\omega)$, thus proving the claim.\\
    \emph{\eqref{item: p local maximizer of mutual information} $\Rightarrow$ \eqref{item: p extreme}:} Let $p$ be no extreme point. Then there is $0 \neq x \in \R^N$ such that $p+x,p-x \in \mathcal{P}$, in particular $x \in Q$. We are going to show that in an arbitrarily close neighborhood of $p$ there is a probability $p' \in \mathcal{P}$ with $\Divergence{f}{p'}{\pind} \geq \Divergence{f}{p}{\pind}$. To this end, let $\epsilon >0$ be arbitrary. Defining $y := \frac{x}{\norm{x}_2} \cdot \frac{\epsilon}{2}$, we see that $p^1 := p+y$ and $p^2 := p-y$ are both in $\mathcal{P}$ and $\frac{\epsilon}{2}$-close to $p$ in the Euclidean norm $\norm{.}_2$. Wlog, assume $\Divergence{f}{p^1}{\pind} \geq \Divergence{f}{p^2}{\pind}$ and set $p' := p^1$. Since $p = \tfrac{1}{2}p^1 + \tfrac{1}{2}p^2$ and $f$ is convex, we find $\Divergence{f}{p}{\pind} \leq \tfrac{1}{2} \Divergence{f}{p^1}{\pind} + \tfrac{1}{2} \Divergence{f}{p^2}{\pind}\leq \Divergence{f}{p^1}{\pind}$, concluding the proof.
\end{proof}

\subsection*{Details of Section \ref{Sub Sec: copulae} and proof of Proposition \ref{Proposition: copula}}

In the following, we write $[\underline{\omega_i}, \overline{\omega_i}]$ by abuse of notation for $ [\underline{\omega_i}, \overline{\omega_i}] \cap \Omega_i$. In particular, $[0,\overline{\omega_i}] = [\min \Omega_i, \overline{\omega_i}] = [\min \Omega_i, \overline{\omega_i}] \cap \Omega_i$. %
The \emph{de-integration function}
\begin{align*}
    \phi \colon \Cop \to Q(\Omega), C \mapsto q^C, q^C(R) := V_C(\vertices(\hat{R})) - V_{\Cind}(\vertices(\hat{R}))
\end{align*}
yields a dependence structure, defined on cuboids $R = [0, \overline{\omega_1}] \times \ldots \times [0, \overline{\omega_n}]$ with vertices $\vertices(\hat{R}) = \{0, \overline{\omega_1}\} \times \ldots \times \{0, \overline{\omega_n}\}$, and extended to all measurable sets on $\Omega$. %
Indeed, any copula $C \in \Cop$, is a multivariate cumulative distribution function, thus defining a probability measure $p^C$ on $\Omega$, by setting $p^C(R) := V_C(\vertices(\hat{R}))$ on cuboids \citep{nelsen2006introduction}. Note that $V_{\Cind}(\vertices(\hat{R})) = \Cind(\omega) = \omega_1 \cdot \ldots \cdot \omega_n =  \pind(R)$. Let $E_i \subseteq \Omega_i$ for some $i \in \{1, \ldots, n\}$. Since intervals generate all Borel sets, we may assume $E_i = [a_i,b_i]$. Note that $[E_i] = E_i \times \Omega_{-i} = [0,b_i] \times \Omega_{-i} \setminus [0,a_i] \times \Omega_{-i}$, so that $p^C([E_i]) = b_i - a_i$ for any $C \in C$. Consequently, $q^C([E_i]) = p^C([E_i]) - p^{\Cind}([E_i]) = 0$, i.e., $\phi(C) \in Q$ and $\phi(C) + \pind \in \mathcal{P}$. Since $V_C$ is linear in $C$, $\phi(\lambda C_1 + (1-\lambda C_2)) = \lambda \phi(C_1) + (1-\lambda) \phi(C_2)$, i.e., $\phi$ respects convex combinations.

Consider now the \emph{integration function} $\psi$ that evaluates the measure $\pind+q$ on a cuboid with corner in the origin, i.e.,
\begin{align*}
    \psi &\colon \phi(\Cop) \to \Cop, q \mapsto C^q, C^q(\omega_1, \ldots, \omega_n) := (\pind + q)([0, {\omega_n}] \times \ldots \times [0, {\omega_n}]).
\end{align*}
Note that $\psi$ converts a (signed) measure (back) into its distribution function by integrating the cuboid from the origin to the vertex $(\omega_1, \ldots, \omega_n)$. This process is also linear and inverts $\phi$.

\end{document}